\documentstyle[epsfig]{mn}

\newcommand{\mnras}{MNRAS}
\newcommand{\apj}{ApJ}

\title[The subhalo populations of $\Lambda$CDM dark haloes]
{The subhalo populations of $\Lambda$CDM dark haloes}

\author[L.~Gao, et al.]
        {L.~Gao, $^1$\thanks{Email: gaoliang@mpa-garching.mpg.de}
        S.~D.~M.~White $^1$,
        A.~Jenkins $^2$,
        F.~Stoehr $^3$,
        V.~Springel$^1$
        \\
        $^1$Max--Planck--Institut f\"ur Astrophysik, D-85748 Garching,
        Germany \\
        $^2$Institute for Computational Cosmology, Department of
        Physics,University of Durham,South Road, Durham  DH1
        3LE, U.K. \\
        $^3$Institut d'Astrophysique de Paris, 98bis Bd Arago, 75014
        Paris,France
        }
\begin{document}
\label{firstpage} \maketitle
\title{The subhalo populations of $\Lambda$CDM dark haloes}
\begin{abstract}
We investigate the subhalo populations of dark matter haloes in
the concordance $\Lambda$CDM cosmology. We use a large
cosmological simulation and a variety of high resolution
resimulations of individual cluster and galaxy haloes to study the
systematics of subhalo populations over ranges of 1000 in halo
mass and 1000 in the ratio of subhalo to parent halo mass. The
subhalo populations of different haloes are not scaled copies of
each other, but vary systematically with halo properties.  On
average, the amount of substructure increases with halo mass. At
fixed mass, it decreases with halo concentration and with halo
formation redshift. These trends are comparable in size to the
scatter in subhalo abundance between similar haloes. Averaged over
all haloes of given mass, the abundance of low mass subhaloes per
unit parent halo mass is independendent of parent mass.  It is
very similar to the abundance per unit mass of low mass haloes in
the universe as a whole, once differing boundary definitions for
subhaloes and haloes are accounted for. The radial distribution of
subhaloes within their parent haloes is substantially less
centrally concentrated than that of the dark matter. It varies at
most weakly with the mass (or concentration) of the parent halo
and not at all with subhalo mass. It does depend on the criteria
used to define the subhalo population considered. About $90$ per
cent of present-day subhaloes were accreted after $z=1$ and about
$70$ per cent after $z=0.5$. Only about $8$ per cent of the total
mass of all haloes accreted at $z=1$ survives as bound subhaloes at
$z=0$. For haloes accreted at $z=2$, the survival mass fraction is
just 2 per cent. Subhaloes seen near the centre of their parent
typically were accreted earlier and retain less of their original
mass than those seen near the edge. These strong systematics mean
that comparison with galaxies in real clusters is only possible if
the formation of the luminous component is modelled appropriately.
\end{abstract}


\section{Introduction}
According to the standard CDM scenario, structure in our Universe
formed hierarchically. Small-scale fluctuations were the first to
collapse as virialised objects. These then merged to form larger
systems. The inner regions of early virialised objects are very
compact and often survive accretion onto a larger system to become
self-bound subhaloes of their host. Since galaxies form by the
condensation of gas at the centres of early haloes, most cluster
galaxies may well be associated with subhaloes in their host
cluster. Only in recent years have numerical techniques and
computer capabilities advanced to the point where it is possible
to study in detail the properties of such subhaloes (Moore et al.
1998, 1999; Tormen, Diaferio \& Syer 1998; Klypin et al. 1999a,b;
Ghigna et al. 1998, 2000; Springel et al. 2001; Stoehr et al.
2002, 2003). These studies indicate that the `overmerging' problem
in early simulations, i.e. the failure to resolve subhaloes
corresponding to galaxies in cosmological simulations of cluster
haloes, was in part a result of insufficient mass and force
resolution.

Using high resolution resimulations of individual cluster or
galaxy haloes, it is possible to study the properties of subhaloes
in detail. Recent papers by De Lucia et al. (2004), Diemand et al.
(2004) and Gill, Knebe \& Gibson (2004) discuss many aspects of
this topic and present results compatible with but complementary
to those presented below. Most studies to date have been limited
because their analysis has been performed on a small number of
individual haloes. Since halo-to-halo variations are large, this
may prevent the derivation of statistically significant results.
In addition, all studies are still affected at some level by
numerical resolution. The available tests show that the subhaloes
seen in a particular object are reproduced moderately well in
mass, but not in position or velocity, when the same object is
resimulated multiple times with varying resolution (Ghigna et al.
2000; Springel et al. 2001; Stoehr et al 2002, 2003). This is a
result of the well known divergence of neighboring trajectories in
nonlinear dynamical systems.

In this paper, we carry out a systematic study of the properties
of subhaloes in the halo population of a single, large-scale
cosmological simulation, and we complement this by analysing a
multi-resolution set of resimulations of a single `Milky Way'
halo, together with a set of high-resolution resimulations of
eight different rich clusters. These resimulations allow us to
investigate how numerical resolution and halo-to-halo variation
affect the conclusions from our cosmological simulation.  We do
not, however, carry out a full study of the numerical requirements
for fully converged numerical results for the properties of
subhaloes.

Previous studies of subhaloes within haloes of different scale
have emphasised similarities -- to a large extent the internal
structure of a `Milky Way' halo looks like a scaled version of
that of a rich cluster halo (Moore et al. 1999; Helmi \& White
2001; Stoehr et al. 2003; De Lucia et al. 2004; Desai et al.
2004). We show below that this scaling is not exact, and that a
better model assumes the mass distribution of low-mass subhaloes
to be the same as in the Universe as a whole, once the differing
definitions of an object's boundary are accounted for. We show
that galaxy haloes have fewer high-mass subhaloes than rich
clusters because of their earlier formation times. Indeed, even
among haloes of given mass, the number of massive subhaloes
correlates well with formation time, as reflected in the halo's
central concentration.

The emphasis of earlier high resolution work on solving the
`overmerging problem' has given rise to the impression that the
subhaloes are typically objects which formed at very early times.
We demonstrate below that this is not the case. Even at low
subhalo masses, most subhaloes were accreted onto the main halo at
low redshift, in most cases well below $z=1$. This is important
when considering the formation paths of present-day cluster
galaxies.

Our paper is organized as follows. We introduce our various
simulation sets in Section 2. In Section 3, we compare the halo
mass abundance function measured from our cosmological simulation
with theoretical predictions and with earlier numerical data. In
Section 4, we investigate the subhalo population as a function of
halo mass and of redshift. The spatial distribution of subhaloes
within haloes is also discussed in Section 4. In Section 5 we
investigate the infall and mass-loss histories of present-day
subhaloes, as well as the fate of objects that are accreted onto
bigger clusters at early times. We discuss our results and set out
our conclusions in Section 6.

\section{The Simulations}
\subsection{The GIF2 cosmological simulation}
We have carried out a cosmological simulation of a $\Lambda$CDM
universe in a periodic cube of side 110 $h^{-1}$Mpc. The total
number of particles is $400^3$, and the individual particle mass
is $1.73\times10^9h^{-1}{\rm M\odot}$. This is a factor of 8
better than the mass resolution of the GIF simulations published
by Kauffmann et al. (1999) but otherwise the parameters and output
strategy of the simulations are rather similar. We therefore call
our new simulation GIF2. The cosmological parameters adopted are:
$\Omega=0.3$, $\lambda=0.7$, $\sigma_{8}=0.9$, and $h=0.7$; We
choose initial fluctuation power spectrum index $n=1$, with the
transfer function produced by CMBFAST (Seljak \& Zaldarriaga 1996)
for $\Omega_bh^2=0.0196$.

Initial conditions were produced by imposing perturbations on an
initially uniform state represented by a `glass' distribution of
particles. This we generated with the method developed by White
(1993) which involves evolution from a Poisson distribution with
the sign of Newton's constant changed when calculating peculiar
gravitational forces. Fluctuations are imposed using the algorithm
described in Efstathiou et al. (1985). Based on the Zeldovich
(1970) approximation, a Gaussian random field is set up by
perturbing the positions of the particles and by assigning them
velocities according to the growing mode solution of linear
theory.

In order to save computational time, we performed the simulation
in two steps. First, we ran the simulation from high redshift
until $z=2.2$ with the parallel {\small SHMEM} version of {\small
HYDRA} (Couchman, Thomas \& Pearce 1995; Macfarland et al. 1998).
At these times the particle distributions are lightly clustered
and thus the {\small P3M} based gravity solver is quite efficient.
We then completed the simulation with a tree-based parallel code,
{\small GADGET} (Springel, Yoshida \& White 2001), which has
better performance in the heavily clustered regime.

Since the two codes adopt different force-softening schemes, it is
necessary to match the force shape at the time we switch from one
code to the other. The softened force becomes Newtonian at a
distance of about $2.3\epsilon$ for {\small HYDRA}, while this
occurs at a distance of $2.8\epsilon$ for {\small GADGET}.
Experimentation showed that a factor of 1.06, namely
$\epsilon_{\rm Hydra}= 1.06\epsilon_{\rm Gadget}$, produces an
excellent match of the two force laws.  In practice, we started
the simulation at $z=49$ with $\epsilon=7h^{-1}$kpc in comoving
units within {\small HYDRA}, and changed the softening to
$\epsilon = 6.604 h^{-1}$kpc for the continuation with {\small
GADGET} after redshift 2.2.

The simulation was carried out on 512 processors of the Cray T3E
at the Rechenzentrum Garching, the supercomputer centre of the
Max-Planck Society. We stored the data at 50 output times
logarithmically spaced between $1+z=20$ and $1+z=1$. This enables
us to construct halo and subhalo merging trees as in Springel et
al (2001). These will be used in other work to model galaxy
formation within the simulation, so that issues of galaxy assembly
and galaxy clustering can be addressed. The numerical data for our
GIF2 simulation are publicly available at
http://www.mpa-garching.mpg.de/Virgo

\subsection{Higher resolution simulations of individual halos}
In order to investigate the importance of numerical and resolution
effects in the study of subhaloes, we have used a set of
multi-resolution resimulations of a Milky Way sized halo carried
out by Stoehr~et al. (2002, 2003). The simulations studied here
are the versions called GA1, GA2 and GA3n in the original papers.
The final mass of the main halo studied here is $M_{200} = 2
\times 10^{12}h^{-1}{\rm M_\odot}$ and its maximum circular
velocity is 240 ${\rm km}s^{-1}$. In this series of resimulations
all perturbation modes present in the initial conditions of a
given resimulation are exactly inherited by all higher resolution
ones. Hence all significant structure in the low resolution
systems should be reproduced at higher resolution. The number of
particles in the high-resolution region, the particle mass and the
gravitational softening are given for the GA simulations in Table
1.

We analyse in addition a set of 8 high-resolution resimulations of rich
cluster halos previously studied in Gao et al. (2004a) and Navarro et
al. (2003).  These simulations all have the same particle mass and
force resolution, $5.12 \times 10^8 h^{-1}{\rm M_\odot}$ and
$\epsilon=5h^{-1}$kpc, respectively.  The clusters were originally chosen as
all objects in a relatively narrow mass range within the $0.479 h^{-1}$Gpc
cosmological simulation of Yoshida et al. (2001). The initial particle number
in the high resolution region of each simulation and the mass of the final
virialized object are listed in Table 2.

All these high-resolution resimulations assumed the same
cosmological parameters as our GIF2 simulation, and all were all
run with the publicly available code Gadget 1.1.

\begin{table}
\caption{Numerical parameters for the GA-series simulations.}
\begin{center}
\begin{tabular}{l c c c c }
   \hline
    & $GA0$  & $GA1$  & $GA2$ & $GA3n$ \\
    \hline
    $N_p$    &68323 &637966 &5953033  &55564205\\
    $m_p[h^{-1}{\rm M_\odot}]$ & $1.8\times 10^{8}$  & $1.9\times 10^{7}$ & $2.0\times 10^{6}$ &
$2.5\times 10^{5}$ \\
    $\epsilon[h^{-1}{\rm kpc}]$ &  1.8   & 1.0 & 0.48 & 0.24\\
     \hline
\end{tabular}
\end{center}
\end{table}

\begin{table*}
\caption{Particle number in the high resolution region and final
$M_{200}$ for the 8 cluster simulations.}
\begin{center}
\begin{tabular}{l c c c c c c c c}
\hline
        & $C1$ & $C2$ & $C3$ & $C4$ &$C5$ &$C6$ &$C7$ &$C8$ \\
\hline
      $N_p$ &8457516 &7808951 &13466254 &9352943 &9011020
              &8704054 &10182210  &8454580 \\
      $m_{200}[h^{-1}{\rm M_\odot}]$ & $0.81\times 10^{15}$  & $0.75\times 10^{15}$
& $0.52\times 10^{15}$ & $0.54 \times 10^{15}$ &$0.62\times
10^{15}$
& $0.84\times 10^{15}$ & $0.45\times 10^{15}$ & $0.60\times 10^{15}$ \\
\hline
\end{tabular}
\end{center}
\end{table*}

\section{The mass function of haloes}

We have used a friends-of-friends group-finding algorithm (Davis
et al. 1985) with the standard linking length of 0.2 in units of
the mean interparticle separation to identify virialised haloes in
our GIF2 simulation. Only haloes which contain at least 20
particles are included in the halo catalogues we analyse below.

The halo mass function (the abundance of haloes as a function of
their mass) is one of the fundamental quantities characterising
the nonlinear distribution of mass in the Universe. Substantial
effort has gone into building theoretical models for this function
and into calibrating them with numerical simulations (Press \&
Schechter 1974; Bond et al. 1991; Lacey \& Cole 1993, 1994; Mo \&
White 1996; Sheth \& Tormen 1999; Sheth, Mo \& Tormen 2001;
Jenkins et at. 2001; Reed et al. 2003; Yahagi et al. 2004). Here we
use our GIF2 simulation, which has a reasonable volume and good mass
resolution, to compare the {\small FOF} halo mass distribution
against published fitting formulae for halo masses down to
$4\times 10^{10}h^{-1}{\rm M_\odot}$ and for redshifts up to
$z=5$.

\begin{figure}
\vspace{-1cm}
\centerline{\psfig{figure=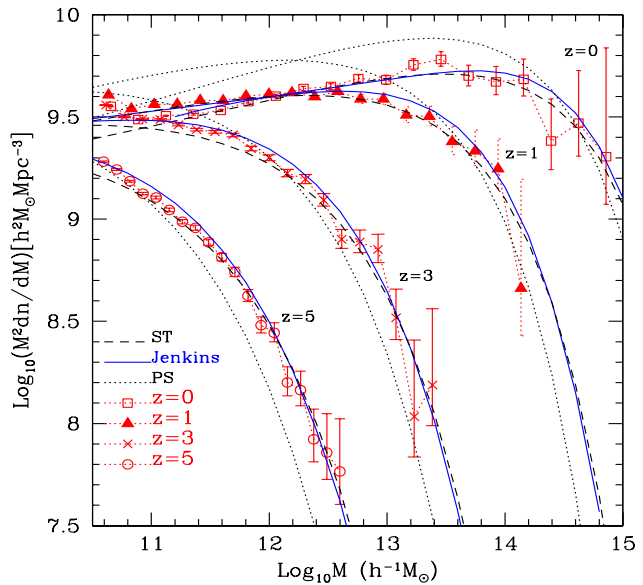,width=250pt,height=250pt}}
\caption{Comparison of the differential halo mass function in our
GIF2 simulation with different analytic predictions. Halos were
identified with a standard {\small FOF} algorithm with linking
length $b=0.2$, and we plot data for all haloes containing more
than 20 particles. Note that we have multiplied the mass function
by $M^2$ to take out the dominant mass dependence.} \label{mdf}
\end{figure}

In Fig.~\ref{mdf}, we plot the differential halo mass function
measured directly from the GIF2 simulation (red dotted line), the
theoretical predictions from Press-Schechter theory (dotted line)
and from Sheth \& Tormen(1999)(dashed line), and the fit to
numerical data published by Jenkins et al. (solid line). Note that
we plot the mass function of Jenkins et al. only over the mass
range where their fitting formula was checked. We have multiplied
the mass function by $M^2$ before plotting in order to take out
the dominant mass dependence and to make the differences between
the various formulae more apparent. Fig. 1 clearly shows that, in
the redshift and mass range studied, the {\small FOF}(0.2) halo
mass function is well described by the formulae of Jenkins et al.
and of Sheth \& Tormen. While being not perfect, the fit is
extremely good in comparison  with the Press \& Schechter mass
function. This confirms the recent conclusion of Reed et al.
(2003) and Yahagi et al. (2004), based on simulations of smaller
volumes, that these formulae can be applied at earlier redshift
and to lower masses than previously demonstrated.

\begin{figure}
\resizebox{8.0cm}{!}{\includegraphics{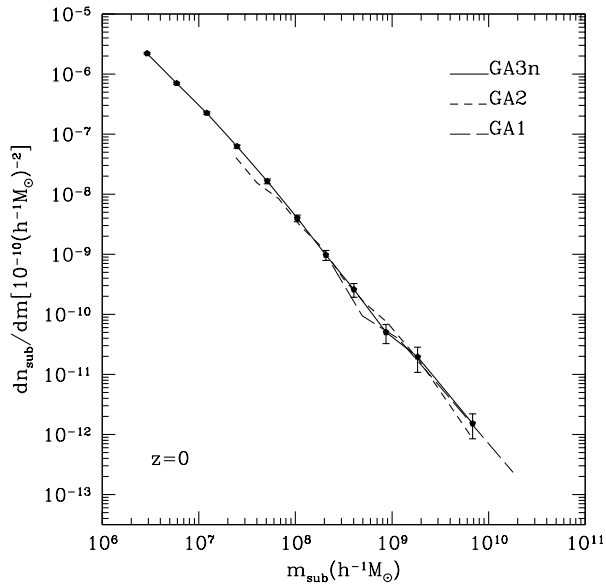}}
\caption{Differential subhalo abundance functions per unit host
mass for the final haloes in our GA1, GA2 and GA3n simulations.
Error bars assume Poisson uncertainties in the counts.}
\label{gamdf}
\end{figure}

\section{Subhalo populations}
Several methods have been proposed to identify subhaloes within
larger systems. For a detailed review we refer to Springel et al.
(2001; hereafter SWTK). In this paper, we use the algorithm
{\small SUBFIND}, developed by SWTK, to isolate locally overdense
and self-bound particle sets within dark matter haloes.  All such
subhaloes containing at least 10 particles are included in our
subhalo catalogues.

\subsection{A convergence study of subhalo populations}
Independent of the particular method employed to identify
subhaloes, most published studies agree that the differential
subhalo mass function (MDF) of an individual halo is approximately
a power-law, $dn/dm \sim m^{-\alpha}$, with $\alpha=1.7-1.9$
independent of redshift and of the mass of the parent halo
(Moore~et al. 1999; Ghigna et al. 2000; De Lucia et al. 2004).  No
study so far has compared in detail the properties of the
subhaloes identified by different methods. Different criteria for
defining the boundaries and the membership of subhaloes are bound
to lead to systematic differences in subhalo populations, but the
uniformity of the derived slopes suggests that such differences
may be correctable through simple scaling factors.

Further study of the effects of numerical resolution on simulated
subhalo populations is clearly important. Numerical convergence
was claimed by Ghigna et al. (2000; hereafter G00), by SWTK and by
Stoehr et al (2002, 2003) on the basis of multi-resolution
simulations of individual objects. However, the data presented are
not fully convincing. For example, Fig.~5 of SWTK shows the
subhalo mass function for a rich cluster resimulated 4 times with
increasing mass and force resolution.  The subhalo abundance in
the lowest resolution simulation S1 agrees well with that in the
highest resolution simulation S4, while the intermediate
resolution simulations S2 and S3 agree very well with each other
but appear significantly offset from S4.  The reasons for this are
unclear. We show similar data in Fig.~\ref{gamdf} for the subhalo
abundance in the GA series resimulations of a `Milky Way' halo. (A
cumulative version of this plot is given by Stoehr et al. (2002)
but without GA3n data).  Here agreement is excellent for subhaloes
that contain at least 30 particles, but there may be significant
differences for smaller subhaloes. These could be due to
resolution problems. As we show below (Section 4.6 and Fig.~10) it
appears that subhaloes with small $N$ dissolve overly fast,
particularly in the inner regions of a halo.

\begin{figure}
\resizebox{8.0cm}{!}{\includegraphics{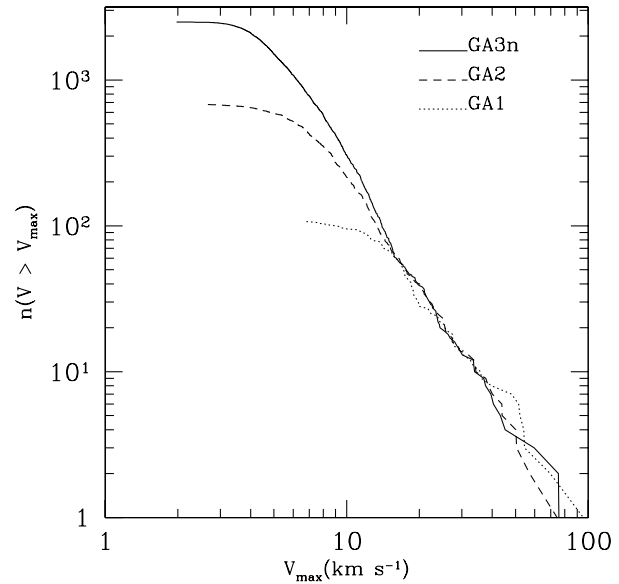}}
\caption{The cumulative abundance of subhaloes as a function of
maximum circular velocity $V_{\rm max}$ for the final haloes in
the GA1, GA2 and GA3n simulations.} \label{gavdf}
\end{figure}

In order to avoid effects due to our particular definition of the
boundary of a subhalo (and so of its mass) we check this
convergence by examining the abundance of haloes in our GA series
as a function of their maximum circular velocity $V_{\rm max}$. We
define the square of this quantity to be the maximum value of
$GM(r)/r$ for those particles identified as bound to the subhalo
by {\small SUBFIND}. $V_{\rm max}$ is a more stable quantity than
the subhalo mass and depends little on how the subhalo is defined.
Fig.~\ref{gavdf} demonstrates that the cumulative abundance of
subhaloes as a function of $V_{\rm max}$ (the VDF) is very well
reproduced between the different simulations in the GA series.
Thus, we conclude that our simulation techniques correctly
reproduce the subhalo abundance down to objects of relatively
small particle count. In particular, GA3n reproduces the correct
subhalo abundance down to values of $V_{\rm max}$ below 10 km/s
and so well below the values relevant to the observed satellites
of the Milky Way.

\begin{figure*}
\centerline{\psfig{figure=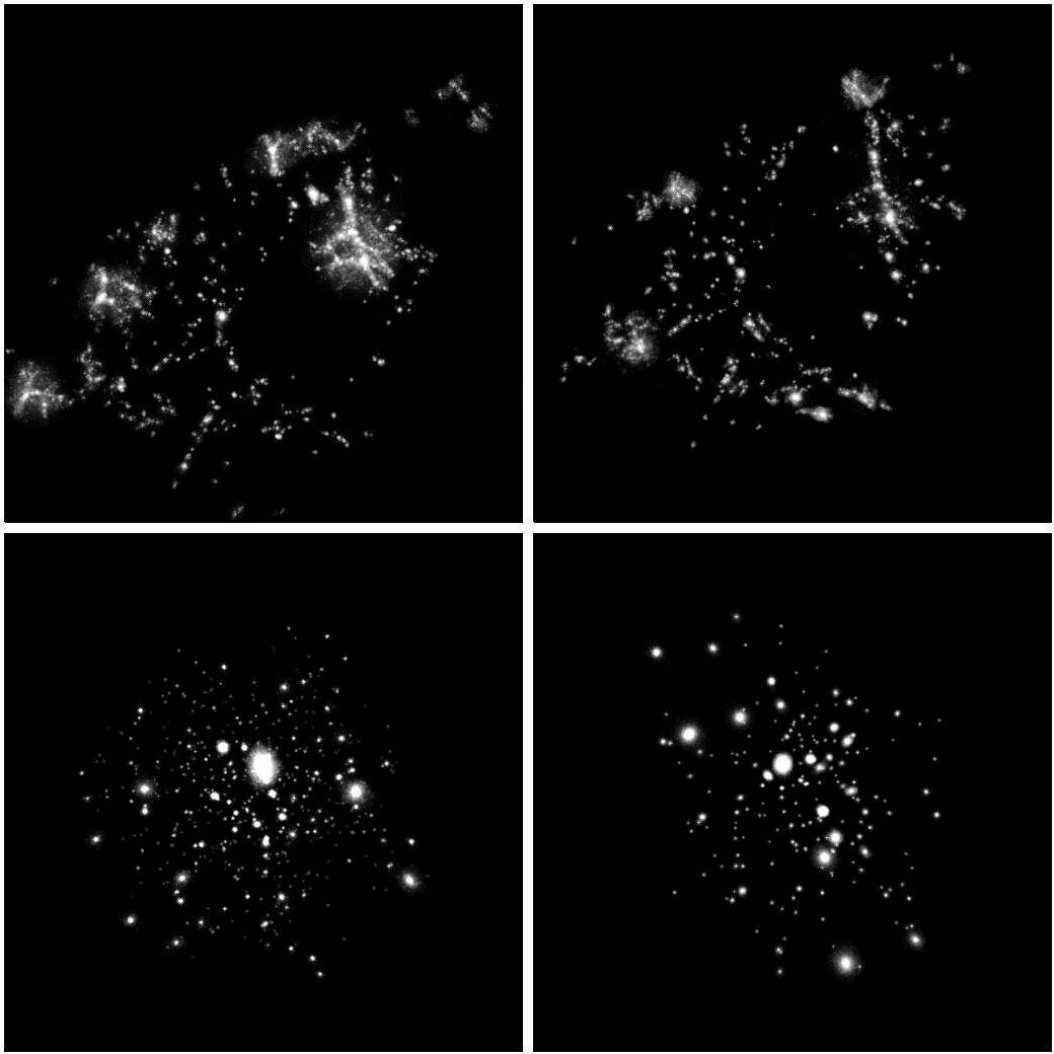,width=400pt,height=400pt}}
\caption{Images at $z=0$ and $z=5$ of the material contained in
$z=0$ subhaloes of the main halo with mass exceeding $5.8 \times
10^8h^{-1}{\rm M_\odot}$ in GA2 and GA3n. Upper plots are for
$z=5$, lower plots for $z=0$. GA2 is shown on the left and GA3n on
the right.} \label{shape}
\end{figure*}

Dark matter haloes are strongly nonlinear and chaotic N-body
systems, so we cannot expect simulations of the `same' object run
with different resolution, with different codes, or with different
integration parameters to be very similar at the final time. (See
for example the various simulations from identical initial
conditions in the Santa Barbara Cluster Comparison Project (Frenk
et al. 1999) This is because in a chaotic N-body system any small
perturbation to the trajectory is amplified exponentially by
subsequent evolution. In the bottom panel of Fig.~\ref{shape}, we
show density maps for all subhaloes belonging to the final {\small
FOF} haloes of GA2 (left-hand panel) and GA3n (right-hand panel).
Although these plots are qualitatively similar, there is no
detailed correspondance between subhaloes. On the other hand, the
upper panels show that the material which makes up these subhaloes
is very similarly distributed in the two simulations at early
epochs. The biggest differences are due to subhaloes which are
included in the final halo in one of the simulations but are just
outside it in the other. Fortunately, we do not care much about
the positions of individual subhaloes and are more interested in
statistical results. A re--simulation of an object with higher
resolution may not reproduce its structure in detail, but it can
still be viewed as the result of evolution from a nearby set of
initial conditions (e.g. Hayes 2003).

\subsection{Is the population of subhaloes similar in all haloes?}
A number of authors have argued that the statistical properties of
subhaloes in a galaxy-sized halo are simply a scaled version of
those in a rich cluster halo (Moore et al. 1999; Helmi \& White
2001; De Lucia et al. 2004; Diemand et al. 2004). {\it Prima
facie} this is surprising, since it is well known that the merging
histories of haloes (and in particular their formation times) vary
systematically with mass (Lacey \& Cole 1993; Navarro, Frenk \&
White 1997). One might expect these differences to result in a
systematic dependence of the subhalo population on mass.

\begin{figure*}
\centerline{\psfig{figure=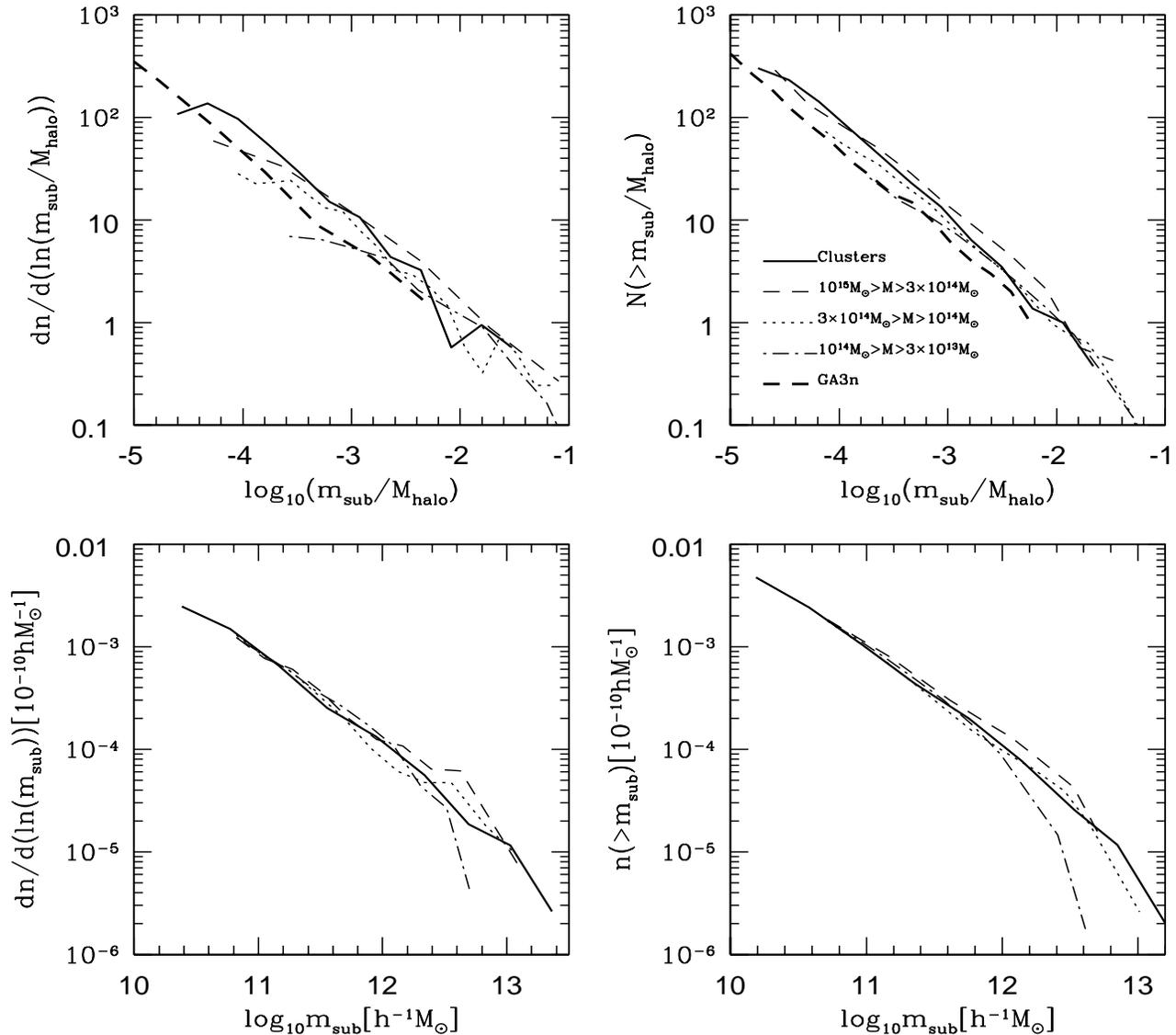,width=500pt,height=500pt}}
\caption{Mass functions at $z=0$ for subhaloes within radius
$r_{200}$ of their parent haloes. In the top left-hand panel we
plot differential subhalo abundance as a function of scaled
subhalo mass, $m_n= m_{\rm sub}/M_{\rm halo}$, for three ranges of
halo mass in our GIF2 simulation, for GA3n and for our 8 cluster
resimulations. In the top right-hand panel, we plot the
corresponding cumulative mass functions. In the bottom left-hand
panel, we plot differential subhalo abundance normalized to the
total mass of the parent haloes, $\langle M_{\rm halo}dn/dm_{\rm
sub}\rangle$. The corresponding cumulative mass functions are
shown in the bottom right-hand panel.} \label{GIF2MF}
\end{figure*}

We define a dimensionless subhalo mass, $m_{n}=m_{sub}/M_{\rm
halo}$, where $M_{\rm halo}$ is the virial mass of the parent halo
defined as spherical region which has 200 times critical density
of universe at that time. In the upper panels of Fig. 5, we plot
subhalo abundance against this normalized mass for three ranges of
halo mass in our GIF2 simulation, $[3\times10^{14}h^{-1}{\rm
M_\odot},\ 10^{15}h^{-1}{\rm M_\odot}]$, $[10^{14}h^{-1}{\rm
M_\odot},\ 3\times 10^{14}h^{-1}{\rm M_\odot}]$ and $[3\times
10^{13}h^{-1}{\rm M_\odot},\ 10^{14}h^{-1}{\rm M_\odot}]$. These
bins contain 7, 33 and 243 haloes, respectively. In this plot we
also include subhalo abundance functions for GA3n and for our 8
cluster simulations. If halo populations of differing mass were
just scaled copies of each other, these various abundance
functions would all agree.  In fact, however, the differential and
cumulative normalized mass functions of Fig. 5 depend
systematically on halo mass. The subhalo abundance in high-mass
haloes is clearly higher (at given {\it scaled} subhalo mass) than
in low-mass haloes. The difference between the rich cluster haloes
and the galaxy halo GA3n is a factor of 2. The cluster haloes also
clearly have more abundant subhaloes than the lowest mass haloes
in our GIF2 simulation. Our simulation data agree with semi-analytical
modelling  by Zentner \& Bullock (2003). These authors argued that,
on average, the subhalo mass fraction should increase with halo mass
because high mass haloes were assembled more recently. A trend in this
direction is also clearly present in in high resolution simulation
data of Diemand et al. (2004), although these authors emphasise the
similarity in subhalo abundance between cluster and galaxy halo rather
than the difference.

In the bottom panels of Fig.~5, we show differential and cumulative
plots of subhalo mass abundance using a different normalization
procedure. We divide the total number of subhaloes in each bin by
the total mass of all the parent haloes to obtain the subhalo
abundance per unit parent halo mass. We then plot this abundance
as a function of the actual mass (rather than the scaled mass).
With this normalization, the subhalo mass functions of different
mass haloes agree very well (see also Kravtsov et al. 2004a). For
relatively low-mass parent haloes the subhalo abundance drops below
that seen in more massive parent haloes for subhalo masses exceeding
about 1 per cent of the parent mass. Ignoring this high mass cut-off,
the subhalo abundance per unit halo mass in Fig.~2 is reasonably well
fit by:

\begin{equation}
dn/dm \simeq 10^{-3.2}(m_{\rm sub} h/{\rm M_\odot})^{-1.9}h{\rm
M_\odot}^{-1}
\end{equation}

An immediate consequence of the universality of this relation is a
shift with parent halo mass in the abundance of subhaloes as a
function of scaled mass $m_n$. For small subhalo masses this shift
is
\begin{equation}
\triangle \log_{10}f(m_n; M_{\rm halo})=0.1\triangle
\log_{10}M_{\rm halo}\, ,
\end{equation}
where $f(m_n; M_{\rm halo})$ is the mean abundance of subhaloes by
normalized mass $dn/dm_n$ in parent haloes of mass $M_{\rm halo}$.
Since the slope of the subhalo MDF is close to 2, this shift in
the normalized function is quite small. As an example, the
abundance shifts by about a factor of 2 at fixed $m_n$ between a
typical galaxy halo of mass $10^{12}h^{-1}{\rm M_\odot}$ and a
rich cluster halo of mass $10^{15}h^{-1}{\rm M_\odot}$. This is
indeed the shift seen between GA3n and the clusters in the upper
panels of Fig.~5

In Fig.~6, we plot the abundance of subhaloes as a function of
$V_{\rm max}$ for GIF2 haloes in our three mass bins and for our 8
clusters. We normalize the abundance as above by dividing the
total subhalo count in each bin by the total mass of the
contributing haloes. This figure confirms the result of Fig.~5.
With this normalization the subhalo abundance as a function of
$V_{\rm max}$ is `universal', i.e. appears not to depend on parent
halo mass. We also plot in Fig.~6 the differential abundance of
{\it haloes} in our GIF2 simulation as a function of $V_{\rm
max}$. Here we normalize by the total mass in the simulation. This
shows the interesting result that subhalo abundance and parent
halo abundance follow similar curves, but with the subhaloes
shifted to lower velocity by 20 or 30 per cent. We will come back to 
this in the next section. Note that the turn-over and drop
at small $V_{\rm max}$ for all these curves are due to the
resolution limit of the simulations.

\begin{figure}
\centerline{\psfig{figure=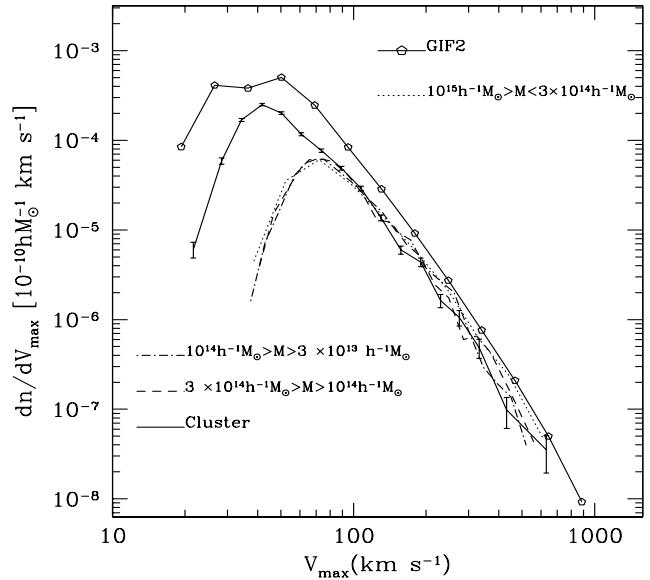,width=250pt,height=250pt}}
\caption{Differential abundance of subhaloes as a function of
maximum circular velocity $V_{\rm max}$. Curves are shown for
three halo mass ranges in the GIF2 simulation and for our 8
cluster simulations. All subhaloes within $r_{200}$ of their hosts
are counted, and the number of subhaloes in each bin is normalized
by the total mass of the contributing haloes. The curve labelled
GIF2 is the corresponding function for the main haloes themselves
and is normalized by the total mass in the GIF2 simulation.}
\label{GIF2vdf}
\end{figure}

\begin{figure}
\centerline{\psfig{figure=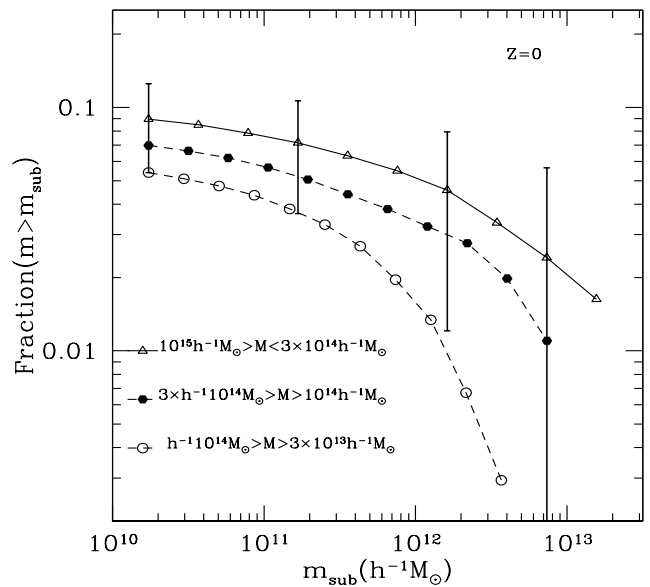,width=250pt,height=250pt}}
\caption{The fraction of halo mass in subhaloes. This plot shows
the fraction of the mass within $r_{200}$ of halo centre which is
in subhaloes more massive than $m_{\rm sub}$ for GIF2 and cluster
haloes in our three mass ranges. Error bars on selected points
show the {\it rms} scatter of the individual values of the mean
for the 15 haloes used to derive the curve for the most massive
bin.}
\end{figure}

\subsection{The mass fraction in subhaloes}

\begin{figure*}
\vspace{-1cm}
\centerline{\psfig{figure=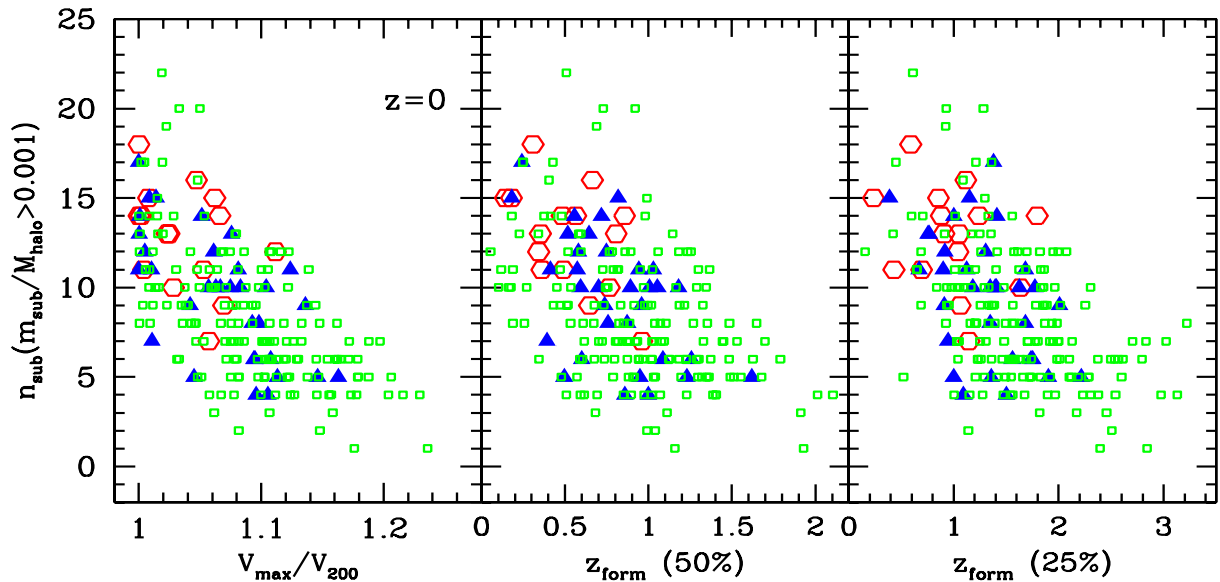,width=500pt,height=500pt}}
\vspace{-8cm} \caption{The relation between subhalo abundance and
the concentration and the formation redshift of haloes. The
left-hand panel shows the number of subhaloes as a function of
halo concentration, as measured by $V_{\rm max}/V_{200}$, for our
GIF2 and cluster simulations. Only subhaloes containing more than
0.1 per cent of the mass of their parent are considered in
compiling these statistics. The middle and right-hand panels show
the same measure of subhalo abundance as a function of halo
formation times defined as the redshifts when the most massive
progenitor has 50 per cent and 25 per cent of the final mass
respectively. Open hexagons are for halos in the mass range $3
\times 10^{14}h^{-1}{\rm M_\odot}<M_{\rm halo} <10^{15}h^{-1}{\rm
M_\odot}$; filled triangles are for halos with $10^{14}h^{-1}{\rm
M_\odot}<M_{\rm halo}<3 \times 10^{14}h^{-1}{\rm M_\odot}$; and
open squares are for haloes with $3 \times 10^{13}h^{-1}{\rm
M_\odot}<M_{\rm halo}<10^{14}h^{-1}{\rm M_\odot}$.}
\end{figure*}

The total fraction of a halo's mass invested in subhaloes is an
interesting quantity but one for which there is little agreement
among the numbers reported in the literature (see, for example,
Ghigna et al. 1998, 2000; Springel et al. 2001; Stoehr et al.
2003). Most authors estimate mass fractions between 5 per cent and
20 per cent, but Moore et al.(2001) argue that the true fraction
might approach unity if subhaloes could be identified down to
extremely small masses. Fig.~7 shows the average mass fraction
(within $r_{200}$) in subhaloes more massive than given $m_{sub}$
for GIF2 and cluster haloes in our three ranges of halo mass.
These curves show clear trends which can already be inferred from
Fig.~5. The subhalo mass fractions appear to converge to
well-defined values as the lower limit on subhalo mass is reduced,
and the asymptotic value is larger for high-mass than for low-mass
haloes. Convergence is a result of the effective slope of the
differential abundance function being larger than $-2$, while the
trend with halo mass results from the apparent universality of the
abundance function at low masses (when normalized by halo mass)
together with a dependence of the high-mass cut-off on halo mass.

The masses of individual subhaloes, and so the value of this
asymptotic mass fraction, will depend systematically on the
algorithm used to define the subhaloes. A variety of different
subhalo identification schemes have been used in published studies
and undoubtedly account in part for the wide range of subhalo mass
fractions quoted. Notice also that since most of the subhalo mass
is in the biggest objects, there is a large halo-to-halo variation
(well over a factor of 2) in the overall subhalo mass fraction. We
show this scatter through the error bars on selected points in the
curve for the most massive haloes in Fig.~7. These give the {\it
rms} scatter of the individual values for the 15 clusters averaged
together to make this curve.

\subsection{Dependence of subhalo populations on halo concentration and
formation time}

As demonstrated in Fig.~5, subhaloes tend to be more abundant in
more massive haloes. In this section, we show that strong trends
are also apparent with halo concentration and with halo formation
time. Such systematics are not surprising since Navarro, Frenk \&
White (1996, 1997) showed that more massive haloes form later and
have lower concentrations. They demonstrated that the density
profiles of CDM haloes are well described by a simple fitting
function with two parameters, $\rho_s$ and $r_s$. Here $r_s$ is a
characteristic radius where the logarithmic density profile slope
is $-2$, and $\rho_s$ is the mass density at $r_s$. They also
showed that these two quantities are strongly correlated, implying
a relation between concentration parameter $c=r_{200}/r_s$ and
halo mass. More massive haloes are less concentrated. They argued
that this is because more massive haloes typically form later.
They also showed that at given mass, haloes which form earlier
have higher concentrations, a result which has been confirmed by
subsequent studies (Wechsler~et al. 2000; Bullock~et al. 2001;
Zhao~et al.  2003a, 2003b). This suggests that haloes of similar
concentration or formation time should have similar formation
histories and so similar numbers of subhaloes.

In the left-hand panel of Fig.~8 we show the number of subhaloes
as a function of the concentration of the host, as measured by
$V_{\rm max}/V_{200}$. (Using this measure of halo concentration
avoids fitting a model to our numerical data). For this
comparison, we count only subhaloes with $m_{\rm sub}/M_{\rm halo}
> 0.001$. This ensures that our results are free of resolution
effects. We include data for our GIF2 haloes and for our 8 cluster
simulations. Haloes of different mass are plotted using different
symbols. Clearly, there is a trend for more concentrated clusters
to contain fewer subhaloes and this trend is present and is
similar in all three mass ranges.

The middle panel of Fig.~8 shows subhalo abundance as a function of
halo formation redshift, defined here as the redshift at which the
most massive progenitor of a $z=0$ halo first exceeds half the
mass of the final object. We obtain this value by linear
interpolation between the redshifts at which we have stored values
of the progenitor masses. In this plot also there is a clear
trend.  Haloes which form late tend to have more subhaloes than
haloes which form early, and the relation between substructure and
formation time is similar for haloes of different mass. Notice
that some haloes form at low redshift yet still contain few
subhaloes.  Examination of some specific cases suggests that these
are products of recent mergers between isolated, similar mass
haloes which had previously eliminated much of their substructure.
In order to avoid such cases, the right-hand panel of Fig.~8 plots
subhalo abundance against a formation time defined as the redshift
when the most massive progenitor has 25 per cent of the final
mass. The number of recently formed objects with little
substructure is reduced and the relation between substructure and
formation time appears cleaner.

\begin{figure*}
\centerline{\psfig{figure=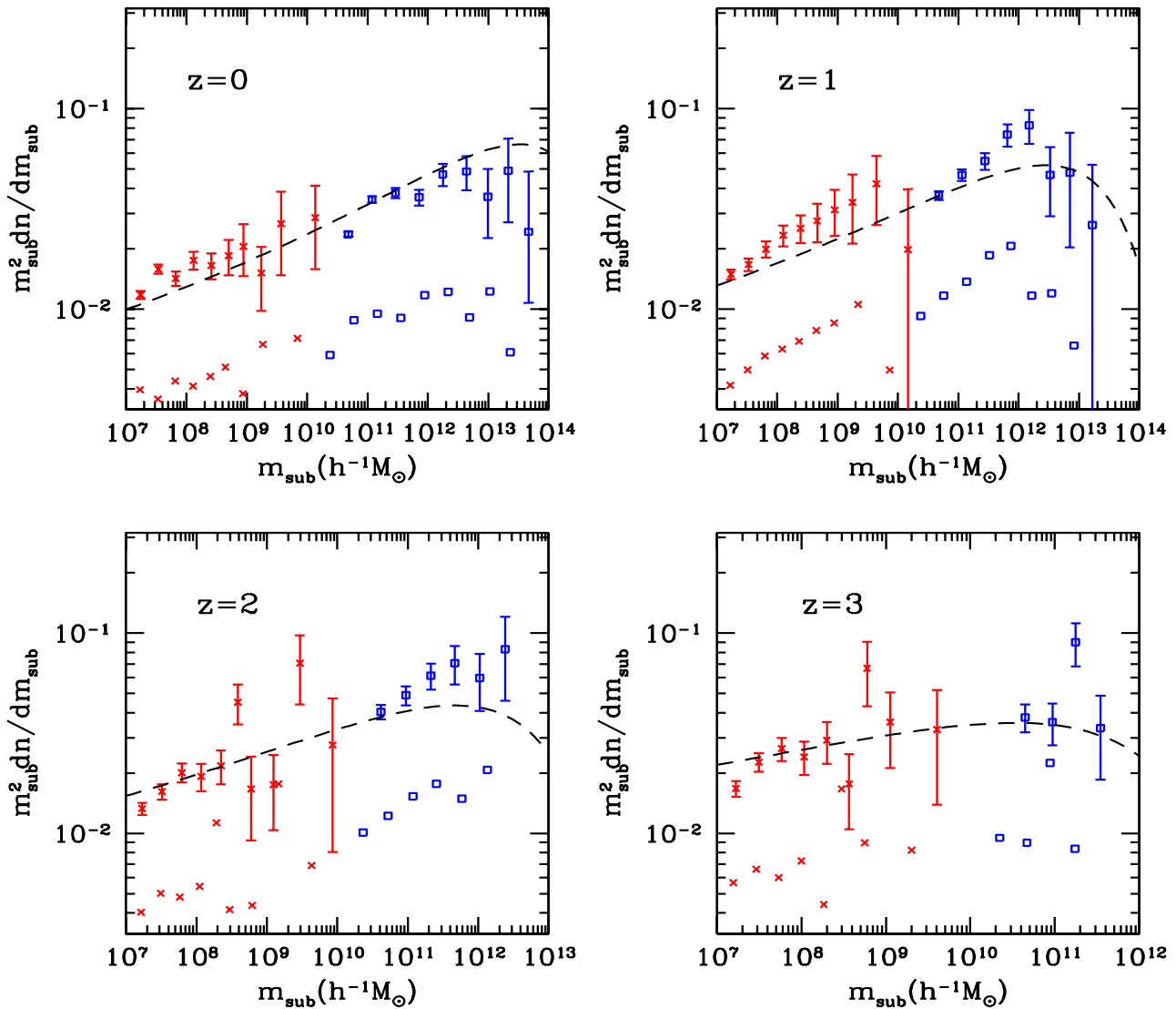,width=500pt,height=500pt}}
\caption{The differential abundance of subhaloes per unit parent
halo mass in the GA3n simulation and in our eight cluster
simulations is compared with the Sheth \& Tormen (1999) formula
for the abundance of haloes per unit mass in the Universe as a
whole.  The four panels refer to four different redshifts as
shown. The simulation results are plotted twice in each panel. The
symbols without error bars are for subhalo masses as returned by
{\small SUBFIND}. The points with error bars are obtained when
these masses are corrected upwards by a factor of two (see text).
The crosses are for GA3n halo; and the squares are the averaged
value of $8$ clusters}
\end{figure*}

A final point to note from Fig.~8 is the scatter in the number of
subhaloes within objects of given concentration or formation time.
The values span a range of up to a factor of four, and the scatter
is at most weakly related to halo mass. Clearly the variety of
possible formation paths for haloes of given global properties is
large enough to produce widely different subhalo populations even
among rather similar objects.

\subsection{The evolution of the subhalo mass function}

Our analysis so far has concentrated on the subhalo distribution
within our simulated haloes at redshift $z=0$. Although this is
the time when our simulations have the best effective resolution
and so can give information over the widest range of scales, it is
nevertheless interesting to look at other redshifts in order to
investigate the evolution of subhalo properties. Given the near
universality we found above, it seems natural to concentrate on
the variation with redshift of the abundance of subhaloes per unit
parent halo mass, and to compare this with the abundance of haloes
per unit mass in the Universe as a whole. This comparison is made
in Fig.~9 using the abundance of subhaloes in the most massive
progenitor of our `Milky Way' halo in GA3n, and of the main
cluster in each of our eight cluster simulations.  For these plots
we multiply the differential abundance distributions by the square
of the (sub)halo mass in order to remove the dominant variation.
We can then plot results corresponding to a range of fourteen
orders of magnitude in abundance and seven orders of magnitude in
(sub)halo mass. The simulation results are shown twice in these
plots, for reasons discussed below. The halo abundance predicted
for the Universe as a whole by the Sheth \& Tormen (1999) mass
function is shown by a dashed line in each panel.

Fig.~9 shows that subhalo abundance distributions vary rather
little with redshift (see also Kravtsov et al. 2004a). At all redshifts
we find the result already noted above for $z=0$. Normalised to total
available mass, the subhalo abundance within haloes is very similar to
the halo abundance in the Universe as a whole. The offset between the
two (the points without error bars and the dashed lines in Fig.~9) is
almost independent of mass and epoch and is roughly a factor of
four in abundance at fixed mass, corresponding to a factor of two
in mass at fixed abundance. This offset can be ascribed to the
different ways in which we define the limits of haloes and of
subhaloes. Our haloes are bounded by a surface within which the
mean interior density is 200 times the critical value, while our
subhaloes are bounded by the surface where their density drops to
the local value in their host. If the internal density profiles of
subhaloes were exactly similar to those of their hosts, and their
radial distribution within their hosts exactly paralleled that of
the dark matter, then it is easy to see that this difference in
boundary definition would cause the masses of subhaloes to be
about a factor of two smaller, on average than those of `field'
haloes of identical structure. The set of points with error bars
in Fig.~9 shows our simulation results when the subhalo masses
returned by {\small SUBFIND} are doubled to `correct' for this
effect. The agreement with the Sheth-Tormen curves is then
remarkably good.

We note that the density profiles of the small haloes which give
rise to subhaloes are more concentrated than those of the larger
haloes they fall into. In addition, we show in the next section
that the radial distribution of subhaloes is less concentrated
than that of the mass. Both these effects should reduce the
difference between the mass assigned to an isolated halo and that
assigned to the subhalo it turns into. On the other hand,
dynamical processes strip material from a halo once it is
incorporated into a larger system, thereby reducing its mass. As
we demonstrate in Section 5, most subhaloes fell into their host
relatively recently and the amount of stripping is typically quite
modest. The combined effect of all these factors is that once
subhalo masses are doubled, as above, the number of subhaloes per
unit mass within a halo is very similar to the number of small
haloes per unit mass in the surrounding universe and thus in the
material from which the main halo formed.

\subsection{The spatial distribution of subhaloes}

How are subhaloes distributed within their parent halo?
Superficially, this appears closely related to the distribution of
galaxies within clusters, but in fact this relation is complicated
because subhalo masses are much more strongly affected by tidal
stripping than are the luminosities of the galaxies they contain.
As a result the effective total mass-to-light ratio of cluster
galaxies is a strongly increasing function of clustercentric
radius (see Fig.~12 of SWTK).

It is also interesting to ask whether the radial distribution of
subhaloes depends on subhalo mass or on the mass of the parent
halo. We address the latter dependence using haloes from our GIF2
and cluster simulations split into the three mass ranges already
analysed in Section 4.2. For each mass range we compute the mean
fraction by number of all subhaloes within $r_{200}$ that lie
within normalized radius $r/r_{200}$.  These subhalo number
density profiles are shown in the upper left-hand panel of Fig.~10
and are compared with a similarly defined profile for the total
mass. All data are shown for $z=0$ and for subhaloes with $m_{\rm
sub}/M_{\rm halo}> 0.001$ only. We can then get comparable and
reliable results for all three halo mass ranges. It is clear that
the radial distribution of subhaloes is substantially less
concentrated than that of the mass as a whole. There is no
significant dependence detected on parent halo mass over the one
order of magnitude range tested in this panel, but a weak
dependence does appear when we compare with our `Milky Way'
simulation GA3n (see below).

\begin{figure*}
\centerline{\psfig{figure=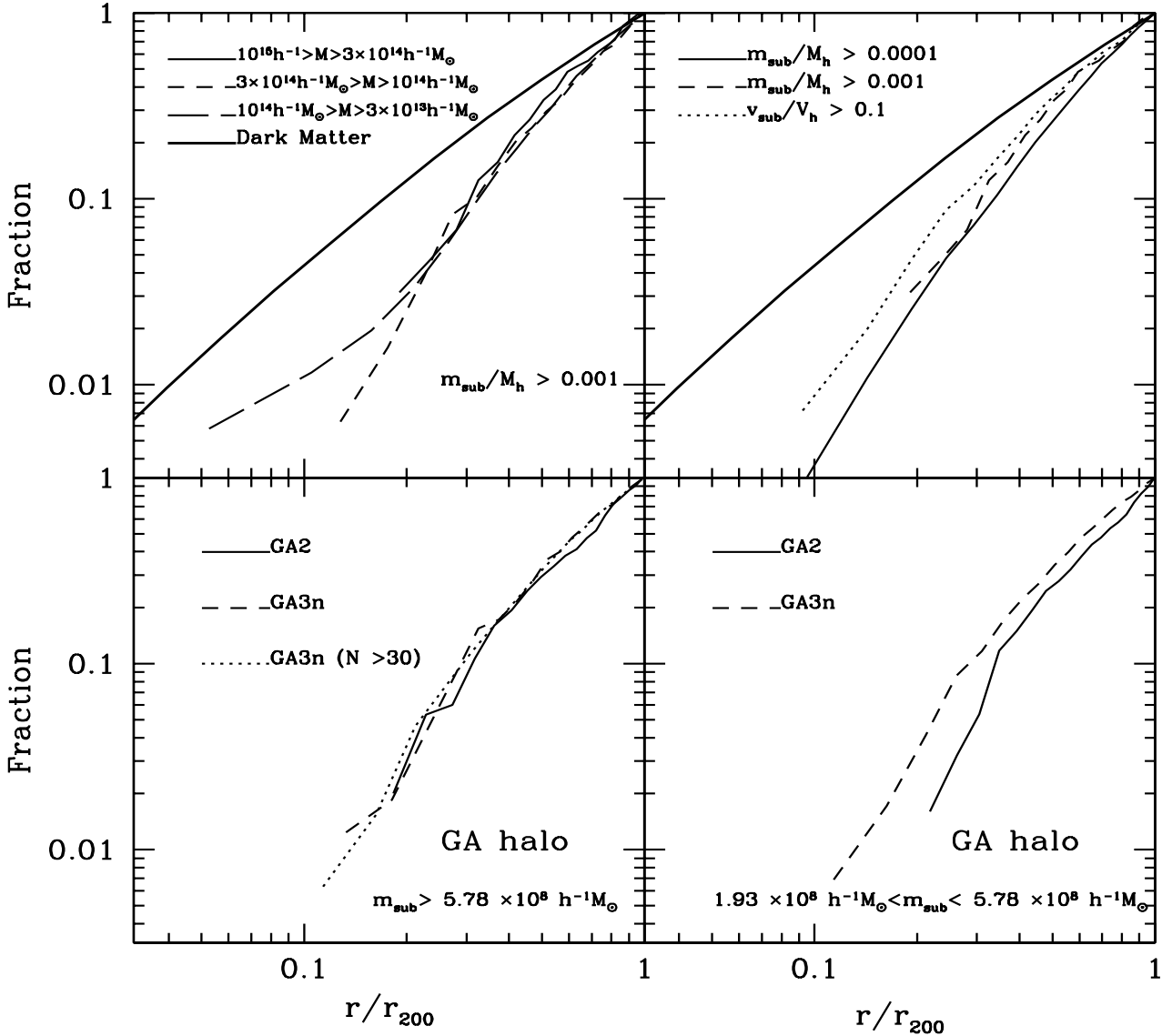,width=500pt,height=500pt}}
\caption{Cumulative radial distributions at $z=0$ for subhaloes
within $r_{200}$ in various sets of haloes in our simulations. The
top left-hand panel shows the fraction of all subhaloes with mass
exceeding 0.1 per cent of their host halo mass and lying within
$r/r_{200}$ of halo centre. Results are plotted for haloes from
our GIF2 and cluster simulations in each of the three mass ranges
discussed above. The top right-hand panel shows similar profiles
but for various subhalo samples of the 15 massive haloes ($M_{\rm
halo}> 3\times 10^{14} h^{-1}{\rm M_\odot}$) in our GIF2 and
cluster simulations. For comparison, we plot cumulative profiles
for the total halo mass in both panels. The bottom left-hand panel
shows profiles for all subhaloes more massive than $5.78 \times
10^8 h^{-1}{\rm M_\odot}$ for two resimulations of a `Milky Way'
halo with mass resolution differing by a factor of 10. This mass
limit corresponds to 30 particles in the lower resolution
simulation. The dotted line shows the profile for subhaloes
containing at least 30 particles in the higher resolution
simulation. The bottom right-hand panel gives subhalo profiles in
these same two simulations but for the subhalo mass range
corresponding to 10 to 30 particles in the lower resolution
simulation.}
\end{figure*}

\begin{figure}
\resizebox{8cm}{!}{\includegraphics{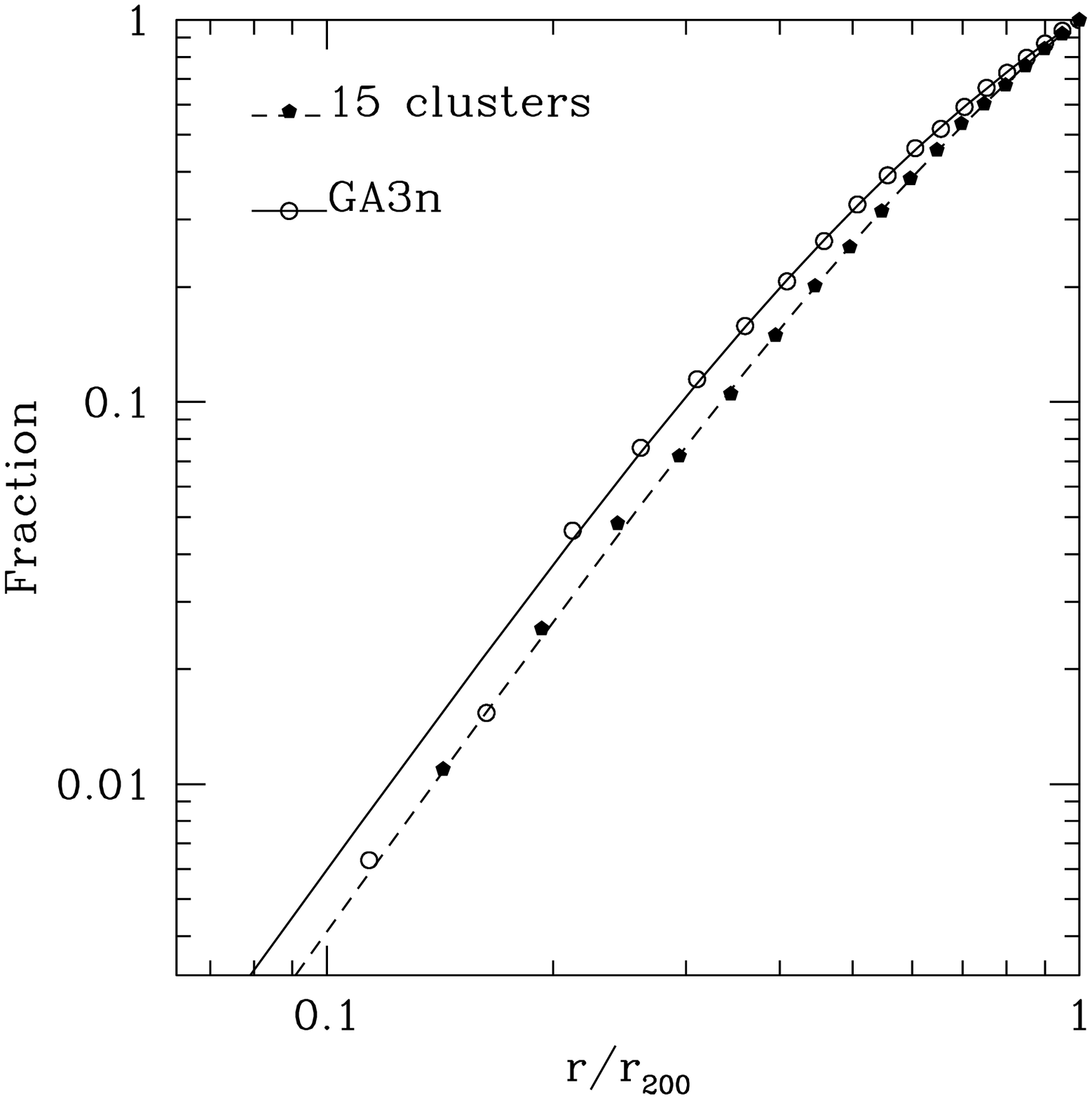}}%
\caption{Cumulative radial distributions at $z=0$ for subhaloes
within $r_{200}$ for the GA3n halo and out 15 clusters. The lines 
overlying the symbols are the corresponding fits given by Equation (3)}
\end{figure}

We address the issue of possible dependences on subhalo mass using
our cluster resimulations together with the haloes in the most
massive bin of our GIF2 simulation (for a total of 15 systems). In
the upper right-hand panel of Fig.~10 we show radial number
fraction plots for subhalo populations limited above $10^{-3}$ and
$10^{-4}$ of the parent halo mass. There appears to be a slight
tendency for the more massive haloes to be more centrally
concentrated, but the effect is small and it is unclear if it is
significant given the relatively small number of parent haloes in
our sample.

For these same 15 clusters, the upper right-hand panel of Fig.~10
also shows the cumulative radial profile of subhaloes for which
$V_{\rm max}$ is greater than 10 per cent of the parent halo's
value of $V_{200}$. It is interesting that this population appears
to be significantly more concentrated than populations defined in
these same haloes above a mass threshold. This presumably results
from a combination of two effects. A subhalo of given density
structure is assigned smaller and smaller masses but largely unchanging
$V_{\rm max}$ values as it gets closer to the centre of its parent
halo. In addition, subhaloes near the centre of their parent tend
to be more heavily affected by tidal stripping than more distant
objects. As demonstrated in Section 5.4, such tidal stripping
affects the masses of subhaloes more strongly than their maximum
circular velocities (Ghigna et al. 2000; Hayashi et al. 2003; 
Kravtsov et al. 2004b).

The lower panels of Fig.~10 use our `Milky Way' simulations to
extend these results to parent haloes of lower mass and to test
further for resolution effects. The dashed and solid curves
compare the cumulative profiles for subhaloes with mass greater
than $5.78 \times 10^7h^{-1}{\rm M_\odot}$ in GA2 and GA3n. This
mass corresponds to 30 particles in GA2 and is $M_{\rm
halo}/40000$. The two profiles agree extremely well, suggesting
that resolution is not seriously effecting our subhalo
distributions.  Reducing the lower limit on subhalo particle
number still further does lead to noticeable effects, as we show
in the lower right-hand panel of Fig.~10. Here the comparison is
repeated for the subhalo mass range corresponding to 10 to 30
particles in GA2.  The abundance of subhaloes is significantly
depressed in the lower resolution simulation, particularly in the
inner regions.  Near the resolution limit of a simulation
subhaloes begin to be lost and they disappear preferentially in
the inner regions of haloes.

Note that the GA3n result in this panel agrees well with that in
the left-hand panel, as does the additional GA3n profile plotted
there for subhaloes with more than 30 particles (and so with
$m_{\rm sub} > 3\times 10^{-6} M_{\rm halo}$). Although all these
profiles are close to those plotted in the upper panels for
mass-limited subhalo populations within haloes of much higher
mass, they are nevertheless noticeably more concentrated. This can
be seen in Fig.~11, where we overplot the 30 particle limited
subhalo number profile of GA3n and the mean profile for subhaloes with
$m_{\rm sub}>10^{-4}M_{\rm halo}$ in our 15 clusters; the subhalo
profiles are plotted with symbols. This suggests that as the
density profile of the parent halo becomes more concentrated, so
too does that of the subhalo population. Note however, that the
effect is much weaker for the subhaloes than for the mass as a
whole. Our subhalo number density profiles are well fit by the
following form:
\begin{equation}\label{pro}
n(<x)/N=(1+ac)x^{\beta}/(1 + acx^{\alpha})
\end{equation}
where, $x$ is the distance to the host centre in units of
$r_{200}$, $n(x$) is the number of subhaloes within $x$, $N$ is
the total number of subhaloes inside $r_{200}$, $a=0.244$,
$\alpha=2$, $\beta=2.75$, and $c=r_s/r_{200}$ is the concentration
of the host halo. The lines in Fig.~11 show the predications of this 
formula for GA3n and for our 15 cluster haloes. Clearly, our
fitting formula works quite well. We caution that the
concentration dependence here is based on our GA-series
simulations only and so should be confirmed with similar
resolution simulations of other objects. We emphasize that this
formula applies to subhalo populations defined above a given lower
mass limit, not to populations defined above circular velocity or
luminosity limits.

Our subhalo number density profiles agree well with those
presented by Diemand et al. (2003) who also found little
dependence either on the mass of the parent or on the mass of the
subhalo. They also agree with the subhalo profiles found by De
Lucia et al. (2004) for their more massive haloes, but not with
the more concentrated profiles found by these authors for their
least massive haloes. The differences are relatively small but
appear significant. In addition, De Lucia et al. (2004) found
massive subhaloes in their simulations to be significantly less
centrally concentrated than low-mass subhaloes. At present, we
have no clear explanation for this difference with our results. We
note that the discrepant results in De Lucia et al are based on a
simulation (denoted M3 by them) which was carried out with an
early version of {\small GADGET} and for which we have other
indications that the chosen integration parameters produced overly
condensed halo cores and thus, perhaps, overly robust subhaloes
(Power et al. 2003). The profiles presented by Gill, Knebe \&
Gibson (2004) are also similar to ours but are somewhat steeper in
the innermost regions. This is likely to reflect the rather
different way in which they find subhaloes and define their
masses.

\begin{figure*}
\epsfysize=16cm \centerline{\epsfbox{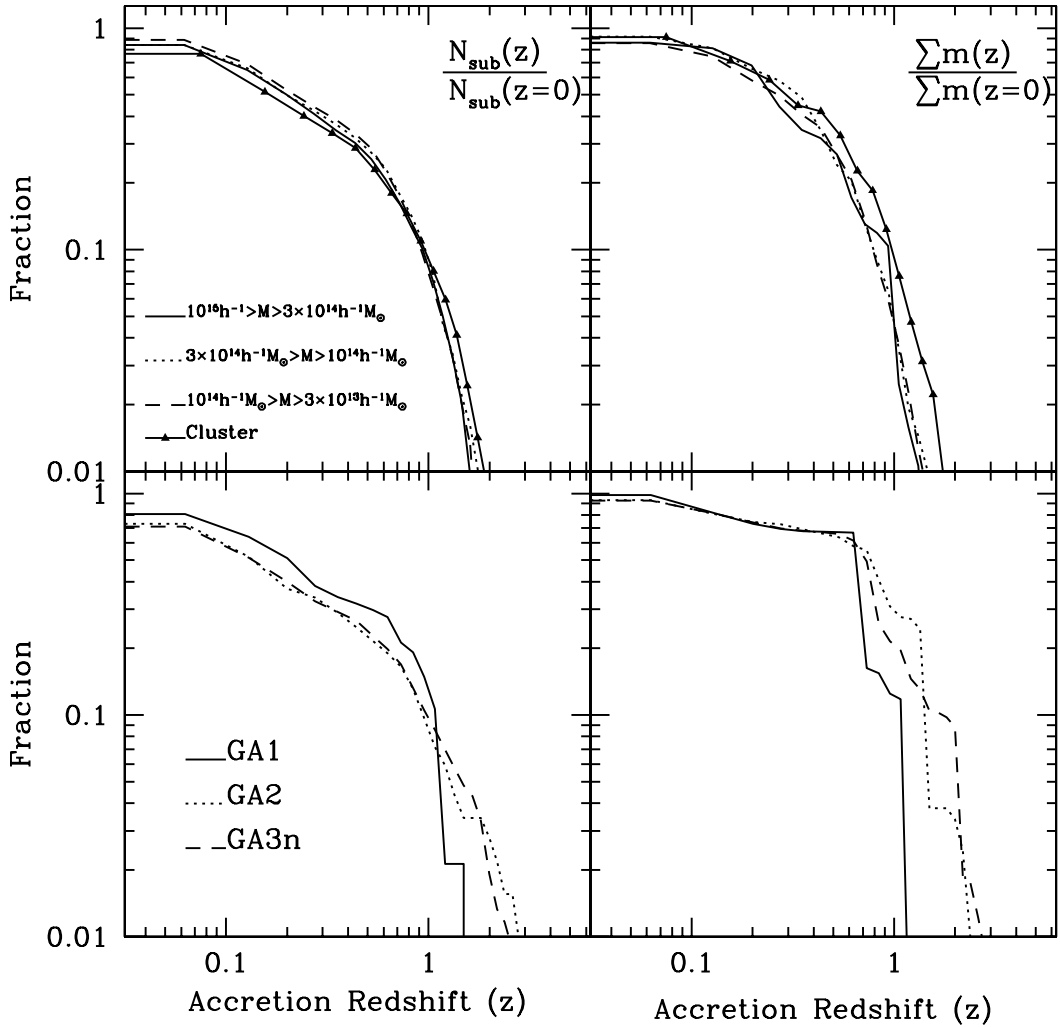}}
\caption{The distribution of subhalo accretion times in our
simulations. The accretion time is defined as the redshift when
the main progenitor first fell into a larger system and so first
became a subhalo. The left-hand panels give the fraction by number
of present-day subhaloes which were accreted before redshift $z$,
while the right-hand panels give the corresponding fractions by
mass. Our different mass halo samples are labelled. The upper
panels refer to our samples of group and cluster haloes, while the
lower panels refer to three simulations of a `Milky Way' halo with
differing mass resolution.}
\end{figure*}

\section{The evolution of subhaloes}
In this section we analyse the evolution of subhaloes by following
the history of individual objects. We construct these histories
according to the definitions of SWTK. Any particular subhalo
identified in one of our stored outputs can have progenitors in
the immediately preceding output which are either subhaloes or
independent {\small FOF} haloes.  A subhalo at the earlier time is
considered a progenitor if more than half its most-bound particles
end up in the subhalo under consideration. A {\small FOF} halo is
considered a progenitor if it contains more than half the
subhalo's particles. The main progenitor of a subhalo is its
largest mass progenitor. By tracing back its main progenitor, the
history of any particular subhalo can be followed to the moment of
`accretion' when its principal halo ancestor fell onto a larger
system and first became a subhalo.

\subsection{The history of present subhaloes}

It is interesting to know when current subhaloes were typically
accreted onto the halo in which they are found. The various panels
of Fig.~12 show, for our different parent halo samples, the
fractions by number (left) and by mass (right) of present-day
subhaloes which were accreted before redshift $z$, as given in the
abscissa. In constructing these plots we have considered all
subhaloes containing at least 10 particles at $z=0$. The group and
cluster mass haloes are shown in the upper panels, and the three
simulations of a `Milky Way' halo are shown in the lower panels.
It is remarkable that very few of the subhaloes identified at
$z=0$ have survived as subhaloes since early times, in agreement with
semi-analytical modelling of Zentner \& Bullock (2003). Only about 10
per cent of them were accreted earlier than redshift 1 and 70 per
cent were accreted at $z<0.5$. These numbers are similar for
haloes of all mass and do not depend significantly on the mass of
the subhaloes considered.  (The apparently discrepant behaviour of
the mass fraction for the GA series is just a consequence of
focussing on a single realisation in which a relatively massive
object happened to accrete at $z\sim 0.7$.) It is clear that
subhaloes are typically recent additions to the haloes in which
they are seen, substantially {\it more} recent, in fact, than
typical dark matter particles.

\begin{figure}
\epsfysize=10.0cm\centerline{\epsfbox{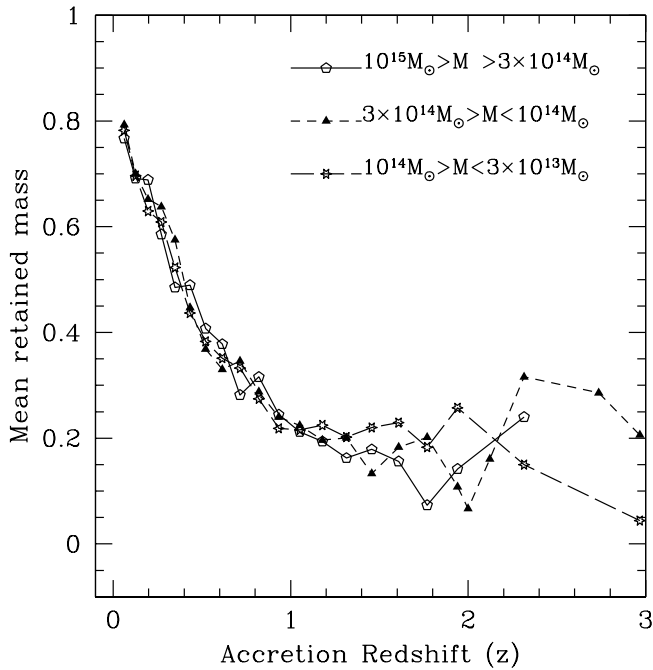}}
\caption{Mean retained mass fractions for subhaloes identified at
$z=0$ as a function of the redshift at which they were accreted.
Different curves refer to parent haloes of different mass and all
present-day subhaloes more massive than $1.73 \times
10^{10}h^{-1}{\rm M_\odot}$ were included when taking the
averages.}
\end{figure}

\begin{figure*}
\epsfysize=16.0cm\centerline{\epsfbox{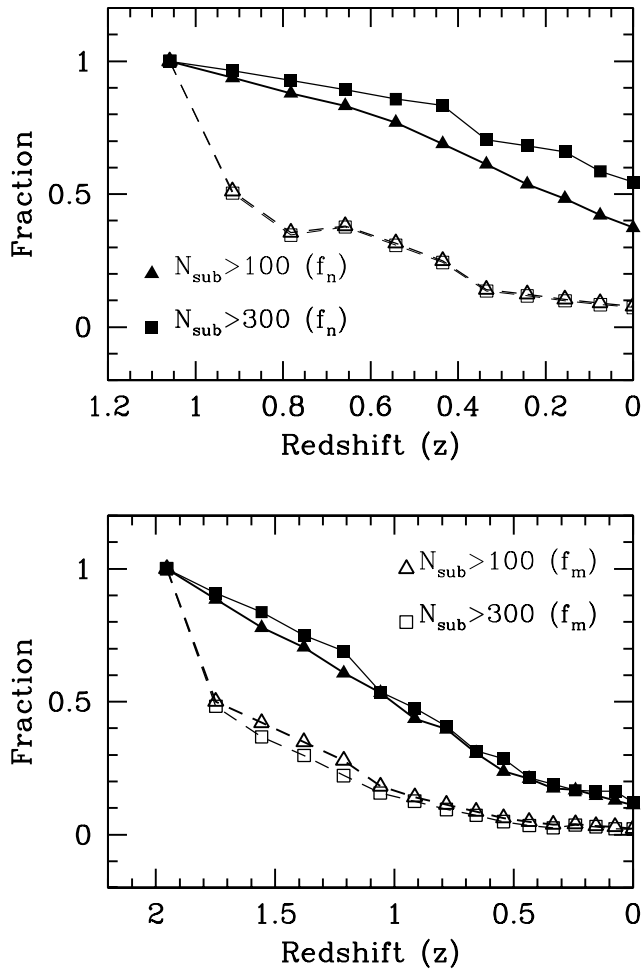}}
\caption{The fate of haloes which merge into the main cluster
progenitor at redshifts of 2 \& 1. Results are shown averaged over
our eight cluster simulations in top and bottom panels,
respectively. Filled symbols and solid lines show the fraction of
haloes which survive as independent subhaloes at each later
redshift $z$, while open symbols and dashed lines show the
fraction of the total progenitor halo mass attached to these
survivors. For each accretion redshift we show results for
progenitor haloes containing at least 100 and at least 300
particles. A surviving subhalo is required to have at least 10
particles assigned to it by our subhalo-finder.}
\end{figure*}
\subsection{Mass loss from subhaloes as a function of time}
When a virialised halo falls onto a bigger structure it loses mass
continually through tidal stripping and its orbit slowly decays
towards the centre of its new parent as a result of dynamical
friction. It is reasonable to expect that subhaloes which fell in
earlier should have lost a larger fraction of their original mass
by the present day.  To measure this mass loss, we calculate the
ratio of the mass of each subhalo at $z=0$ to the mass of its
progenitor halo just before it was accreted.  In Fig.~12 we plot
the mean of this ratio for all present-day subhaloes more massive
than $1.73 \times 10^{10}h^{-1}{\rm M_\odot}$ as a function of
their accretion redshift, showing results separately for parent
haloes of different mass and including haloes from our GIF2 and
cluster simulations. The noise at high redshifts in this plot is
due to poor statistics. As we saw already in the last section,
very few present-day subhaloes were accreted at such early times.

It is clear from Fig.~13 that there is little dependence of mass
loss on parent halo mass and that the mean retained mass fraction
for {\it surviving} subhaloes is a strong function of accretion
redshift. Notice that since we compile statistics for subhaloes
identified at $z=0$, we neglect objects which have been stripped
to masses below our resolution limit or disrupted entirely. As we
show in the next section, the retained mass fractions of Fig.~12
are thus substantially higher than those of typical haloes
accreted at each redshift.

\subsection{The fate of accreted haloes}

In this section we follow all the haloes which are accreted onto
the main progenitor of a final halo (and so first become subhaloes
of it) at redshifts 2 and 1. We are interested to learn what
fraction of these survive until $z=0$, what are the final masses
of the survivors, what happens to those that do not survive, and
how these various fates depend on the mass of the halo which is
accreted. Here we use our eight rich cluster simulations to
investigate these issues.  We begin by finding all progenitors of
a final cluster which are independent {\small FOF} haloes in the
stored output immediately beyond $z=1$ (or 2) but are already
listed as part of the main progenitor in the subsequent output. We
then attempt to trace all these subhaloes forward until either we
reach $z=0$ or they are lost.  Three different fates are possible
for each accreted halo:
\begin{description}
\item[(1)]If it can be followed as a subhalo to $z=0$, we say it survives;
\item[(2)]If it dissolves and becomes part of the main body of its host, we
say it disrupts;
\item[(3)]If it merges with a larger subhalo and then loses its identity, we
say it merges. We find that no more than a few percent of accreted haloes
suffer this fate.
\end{description}

In Fig.~14 we show the fraction of accreted haloes which are
identified as surviving at each later redshift, as well as the
fraction of the total mass initially assigned to these haloes
which remains attached to the surviving subhaloes. We see that
while more than 90 per cent of accreted haloes are identified as
subhaloes in the output immediately after their accretion, the
total subhalo mass is, however, only about half of that assigned
to the original haloes. This is a result of the effect already
noted above. The algorithm which we use to identify subhaloes
bounds them at a substantially higher density than that used to
bound isolated haloes. Consequently, if a field halo falls onto a
larger system its assigned mass decreases by a factor of two, on
average, even if its structure is unchanged. As subhaloes orbit
within their parent haloes, their masses are further reduced by
tidal stripping.  Thus the fraction of the initial mass attached
to the survivors continually decreases, and more and more
subhaloes drop below the mass limit for identifying them in our
simulations.

Fig.~14 gives results for two sets of progenitor haloes at each
accretion redshift. These are defined to contain at least 100 and
at least 300 particles, corresponding to halo masses exceeding $5
\times 10^9h^{-1}{\rm M_\odot}$ and $1.5 \times 10^{10}h^{-1}{\rm
M_\odot}$, respectively. As can be seen, the mass fraction in the
survivors is independent of this mass limit and is 8 per cent for
haloes accreted at $z=1$ and 2 per cent for haloes accreted at
$z=2$. The fraction of survivors by number does depend on the mass
limit. Our samples only contain subhaloes identified with more
than 10 particles, so descendents begin to be lost from the lower
mass halo sample for mass reduction factors greater than 10
whereas factors exceeding 30 are needed to remove objects from the
higher mass sample.

\begin{figure*}
\resizebox{8cm}{!}{\includegraphics{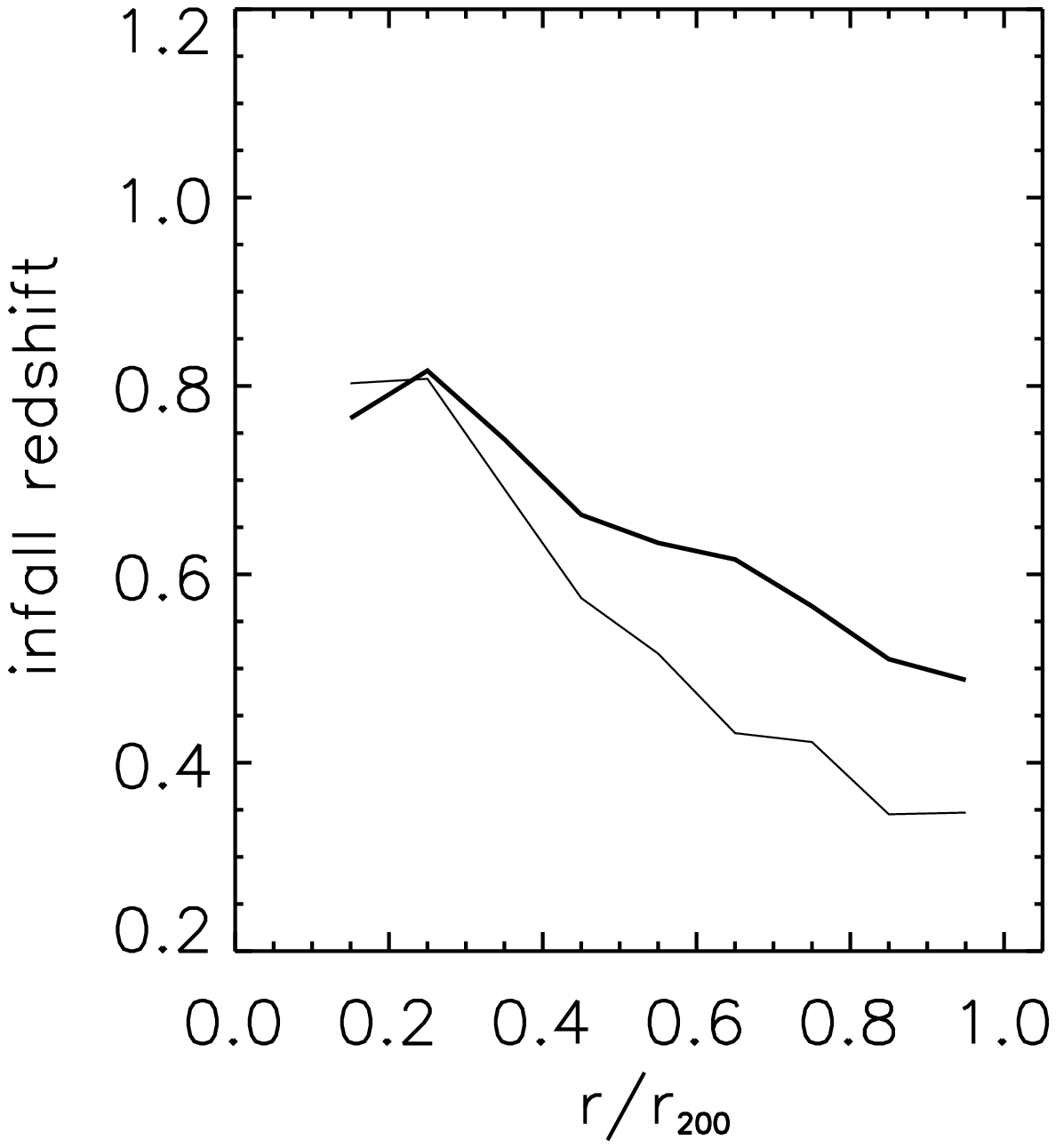}}%
\hspace{0.13cm}\resizebox{8cm}{!}{\includegraphics{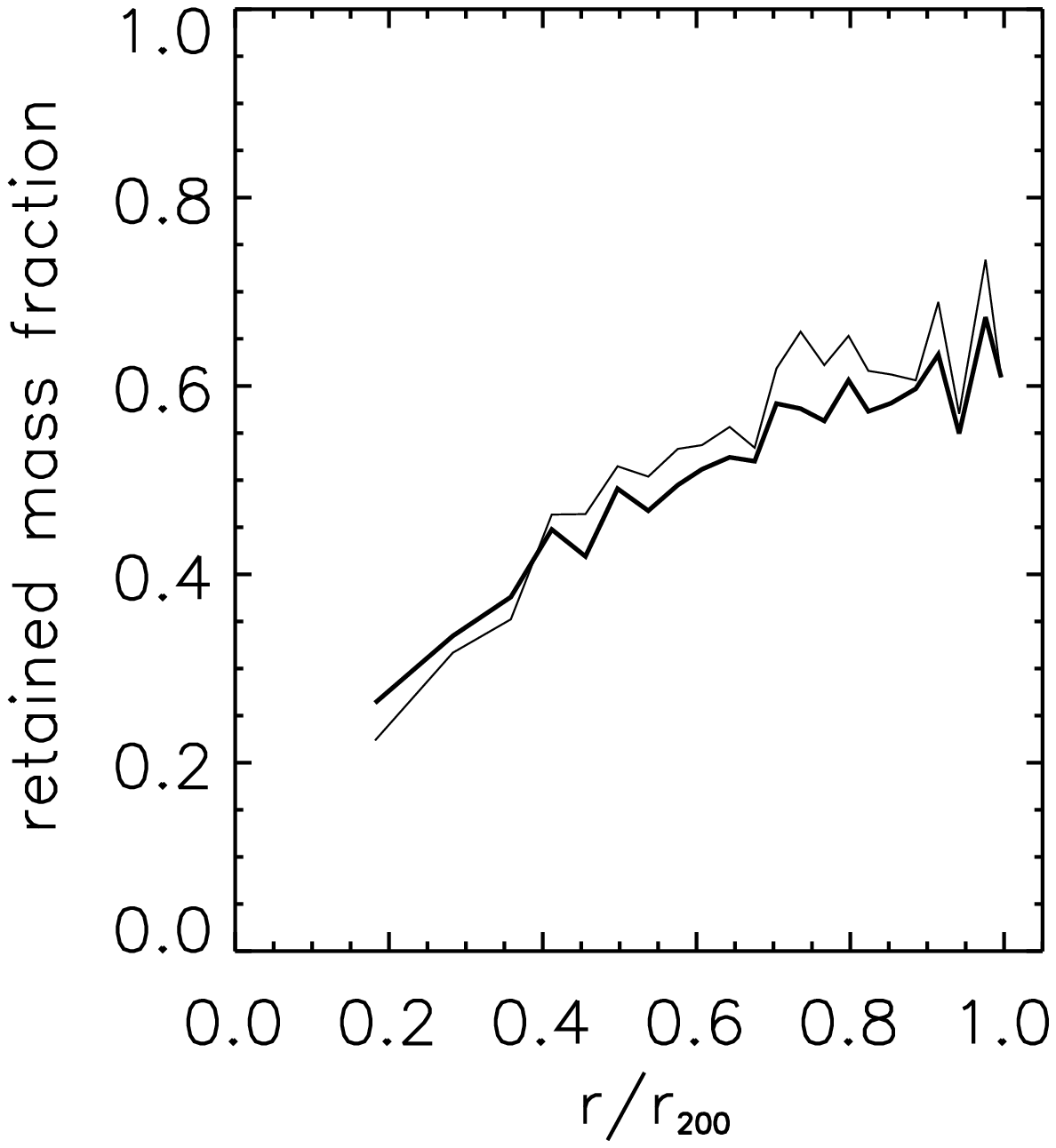}}\\%
\resizebox{8cm}{!}{\includegraphics{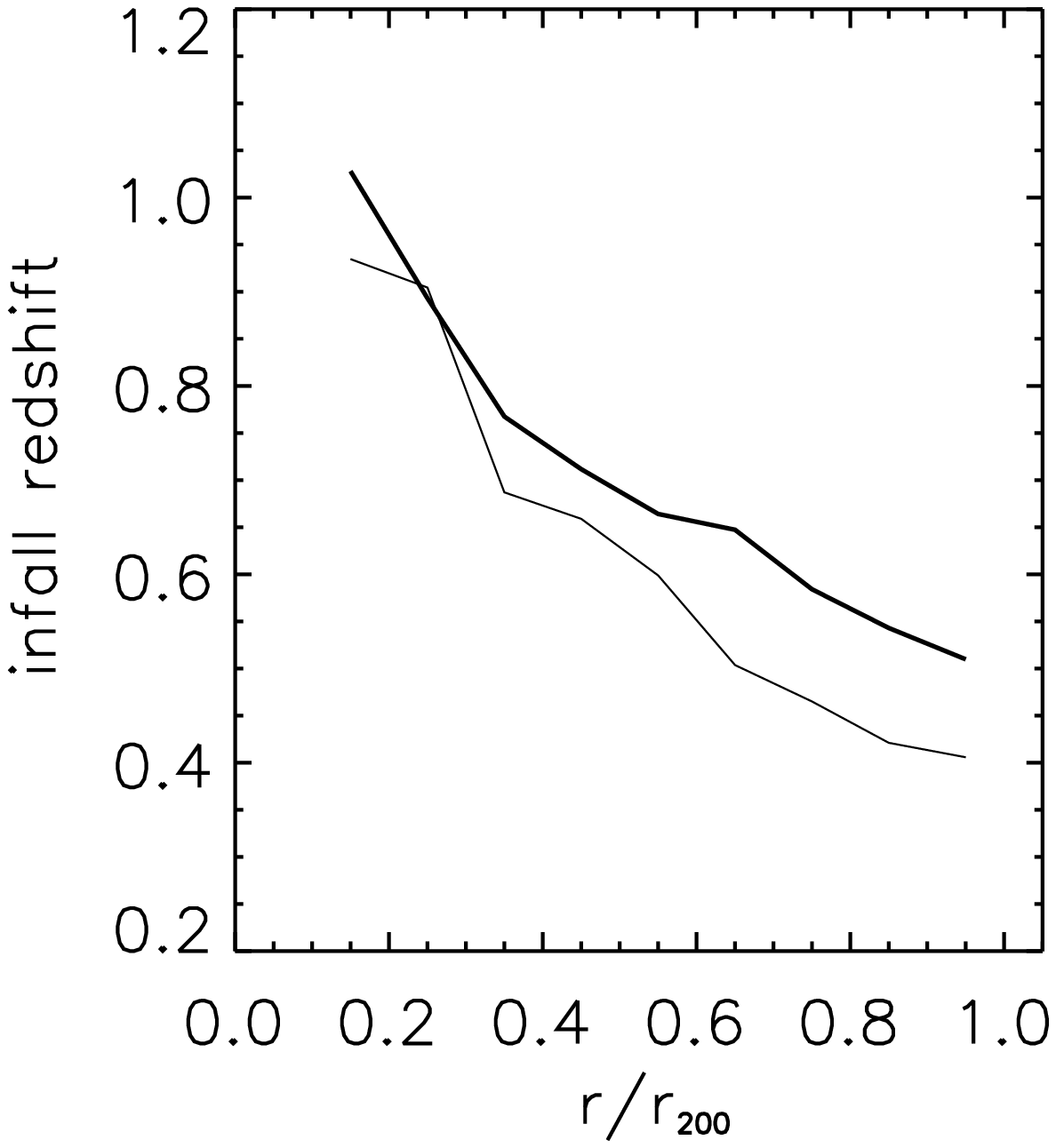}}%
\hspace{0.13cm}\resizebox{8cm}{!}{\includegraphics{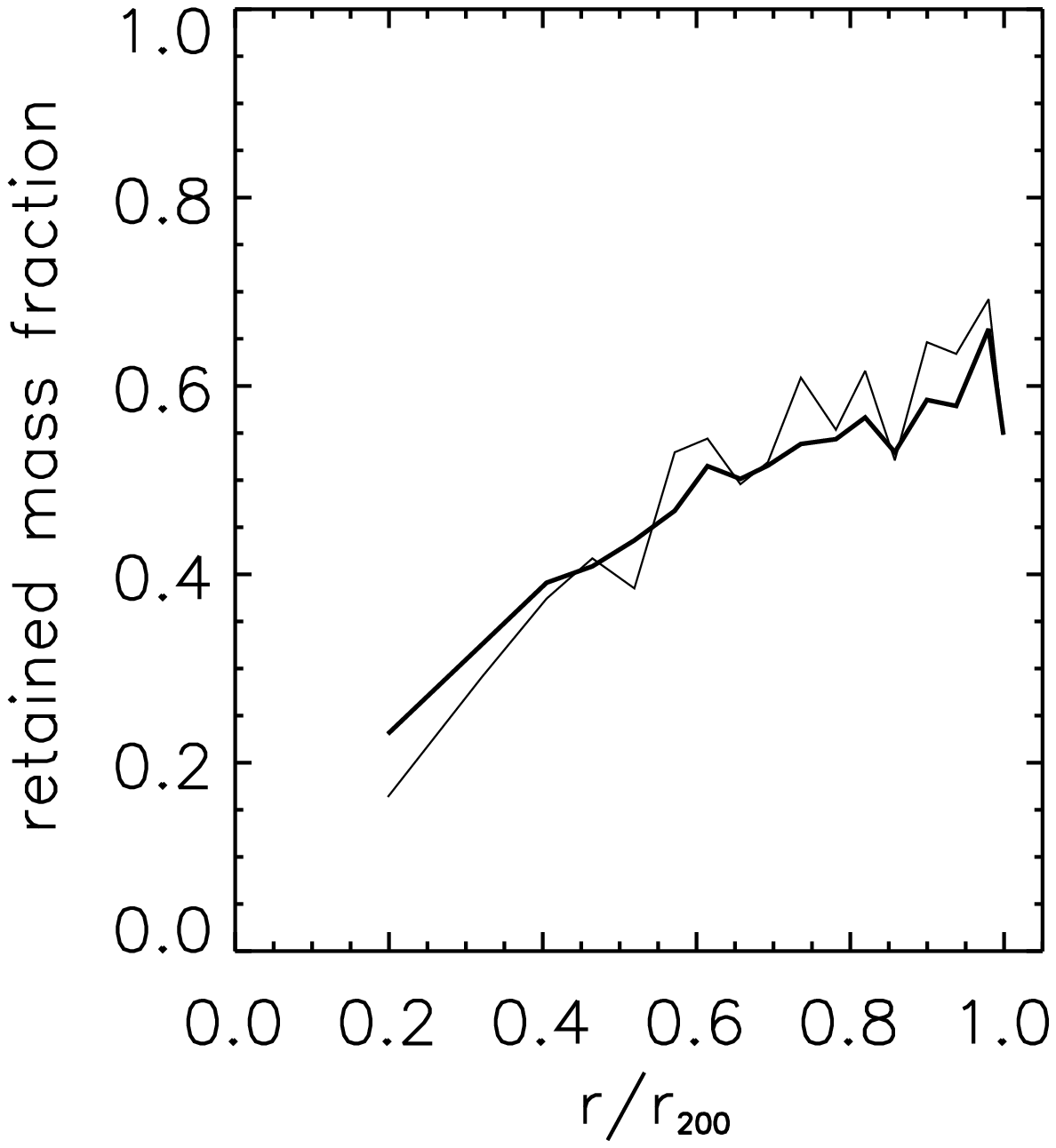}}\\%
\hspace*{1.36cm}\ \\%
\caption{Radial dependence of the accretion redshift (left-hand
panel) and retained mass fraction (right-hand panel) for subhaloes
of the 15 haloes more massive than $3\times 10^{14} h^{-1}{\rm
M_\odot}$ in our GIF2 and cluster simulations. In each panel thick
solid lines give the mean at each value of $r/r_{200}$ while thin
solid lines give the median. The top panels are for subhaloes more
massive than $2\times 10^{10} h^{-1}{\rm M_\odot}$ while bottom
are for subhaloes more massive than $6\times 10^{10} h^{-1}{\rm
M_\odot}$.}
\end{figure*}

\subsection{Radial dependence of accretion time and mass loss}
Subhaloes which were accreted onto their parent halo's main progenitor at
early times initially had relatively short orbital periods and so should be
located, on average, in the inner regions of the final halo. In addition, a
subhalo which has been orbiting within its parent for a long time will have
suffered substantially from the effects of dynamical friction and tidal
stripping, so its orbit will have decayed by a larger factor than that of a
recently accreted subhalo of similar current mass. Both these effects are
expected to lead to a correlation between the radial position of a subhalo and
its accretion redshift.

In Fig.~15 we plot mean and median values of accretion redshift
and of retained mass fraction against $r/r_{200}$ for subhaloes 
of the 15 haloes in our GIF2 and cluster simulations with masses
exceeding $3 \times 10^{14}h^{-1} {\rm M_\odot}$. The upper and l
ower panel refer to subhaloes more massive than $2\times 10^{10}
h^{-1}{\rm M_\odot}$ and more massive than $6\times 10^{10}
h^{-1}{\rm M_\odot}$ respectively. Clearly there is indeed a
strong age-radius relation which is similar for subhaloes of
differing mass. Recently accreted subhaloes tend to occupy the
outer regions of their host, while older subhaloes reside
preferentially in the inner regions. In addition, haloes near the
centre typically retain a much smaller fraction of their
progenitor halo's mass than those in the outer regions (The reasons
for this are discussed in some detail by Kravtsov et al. (2004b)). 
The large difference between the median and the mean in the accretion
redshift plot is a reflection of the substantial skewness of the
distribution. As we already saw in Fig.~12 and 13, tidal stripping 
is clearly very effective and, as a consequence, the ancestors of 
inner subhaloes were more massive than those of outer subhaloes of 
the same mass. Thus in a galaxy cluster inner subhaloes are likely 
to host brighter galaxies than outer subhaloes of similar mass.

\section{Summary and discussion}
We have used a single, large-scale cosmological simulation
together with two sets of resimulations of the formation of
individual cluster and galaxy haloes to carry out a systematic
study of the properties of dark halo substructure in the
concordance $\Lambda$CDM universe. In agreement with the earlier
work of Jenkins et al. (2001), Reed et al. (2003) and Yahagi et
al. (2004) we find the abundance of haloes (defined using a
friends-of-friends group finder with linking length $b=0.2$) to be
well described by the Sheth \& Tormen (1999) mass function down to
masses of a few times $10^{10}{\rm M_\odot}$ and out to a redshift
of 5. Our main results for the subhalo populations within these
haloes can be summarized as follows:

\begin{description}

\item[(1)] The subhalo populations of different haloes are not simply scaled
copies of each other, but vary systematically with global halo properties. On
average, massive haloes contain more subhaloes above any given fraction of
parent mass than do lower mass haloes, and these subhaloes contain a larger
fraction of the parent's mass. At given halo mass, subhaloes are more abundant
in haloes which are less concentrated, or formed more recently.

\item[(2)] There is considerable scatter in the abundance of subhaloes between
haloes of similar mass, concentration or formation time. This presumably
reflects differences in the details of halo assembly.

\item[(3)] For subhalo masses well below that of the parent halo the mean
subhalo abundance {\it per unit parent mass} is independent of the actual mass
of the parent. It is very similar to the abundance of haloes per unit mass in
the universe as a whole, once a correction is made for the differing bounding
density within which the masses of haloes and subhaloes are defined.

\item[(4)] Normalised in this way to total parent halo mass, the mean
abundance of subhaloes as a function of maximum circular velocity
is also quite similar to the abundance per unit mass of haloes as a
function of $V_{\rm max}$. For subhaloes the abundance per unit
mass is about a factor of two lower at given $V_{\rm max}$ than
for haloes. Equivalently, the $V_{\rm max}$ values of subhaloes at
given abundance per unit mass are about 25 per cent lower than
those for haloes.

\item[(5)] In agreement with previous studies, we find the the radial
distribution of subhaloes within their parent haloes to be much
less concentrated than that of the dark matter. We find no
significant dependence of this radial profile on the mass of the
subhaloes and only a very weak dependence on the mass (or
concentration) of the parent halo. To a good approximation the
radial distribution of subhaloes appears `universal' and we give a
fitting formula for it in equation~\ref{pro}.

\item[(6)] The subhalo number density profile does depend on how the
population is defined. Subhalo populations defined above a minimum circular
velocity limit are significantly more concentrated than those defined above a
minimum mass limit.

\item[(7)] Most subhaloes in present-day haloes fell into their
parent systems very recently. Only about 10 per cent of them were
accreted earlier than $z=1$ and 70 per cent were accreted at
$z<0.5$. These fractions depend very little on the mass of the
subhaloes or on that of their parents

\item[(8)] The rate at which tidal effects reduce the mass of
subhaloes is  not strongly dependent on the mass of the accreted
object or on that of the halo it falls into. About 92 per cent of
the total mass of haloes accreted at $z=1$ is removed to become
part of the `smooth' halo component by $z=0$. For haloes which
fall in at $z=2$ this fraction is about 98 per cent. Note that the 
highest mass accreted objects merge into the central regions more
quickly because of dynamical friction effects.

\item[(9)] Subhaloes seen near the centre of their parent haloes typically
fell in earlier and retain a smaller fraction of their original
mass than subhaloes seen near the edge. Thus inner subhaloes may
be expected to host brighter galaxies than outer subhaloes of similar
mass (see Springel et al. 2001). 
\end{description}

These properties suggest a relatively simple picture for the evolution of
subhalo populations. A substantial fraction of the mass of most haloes has
been added at relatively recent redshifts, and this mass is accreted in clumpy
form with a halo mass distribution similar to that of the Universe as a
whole. Since tidal stripping rapidly reduces the mass of subhaloes, the
population at any given mass is dominated by objects which fell in recently
and so had lower mass (and thus more abundant) progenitors. The orbits of
recently accreted objects spend most of their time in the outer halo, so that
subhaloes of given mass are substantially less centrally concentrated than the
dark matter as a whole. Subhaloes which are seen near halo centre have shorter
period orbits and so must have fallen in earlier. They thus retain a
relatively small fraction of their initial mass.

Comparison of these subhalo properties with observation is far
from simple. The recent accretion of most subhaloes means that the
galaxies at their centres were almost fully formed by the time
they became part of their current host. We might therefore expect
their observable properties to be more closely related to the mass
of their progenitor haloes and to their accretion redshifts than
to the current masses of their subhaloes. Explicit tracking of
galaxy formation during the assembly of cluster haloes shows that
these differences can be large. For example, both Diaferio et al.
(2001) and Springel et al. (2001) find radial number density
profiles for magnitude limited samples of galaxies which are
similar both to the underlying dark matter profiles and to the
observed profiles of real clusters, but which are very different from
the number density profiles for mass limited subhalo samples.
Similar differences are to be expected between the velocity biases
of galaxies and subhaloes. Models for the stellar content of
subhaloes which are based purely on their current mass and
internal structure are very unlikely to be successful. The past
history of subhaloes must be included to get realistic results, as
must galaxies associated with apparently disrupted subhaloes. We
investigate these issues further in a companion paper (Gao et al.
2004b);

\section*{Acknowledgements}
The simulations used in this paper were carried out on the Cray
T3E and Regatta supercomputer of the computing centre of the
Max-Planck-Society in Garching, and on the Cosma supercomputer of
the Institute for Computational Cosmology at the University of
Durham. We thanks Yipeng Jing, Chris Power and Xi Kang for  useful
discussions. The numerical data for our GIF2 simulation are
publically available at http://www.mpa-garching.mpg.de/Virgo.
\label{lastpage}

\end{document}